\documentclass[12pt, letterpaper]{article}
\pdfoutput=1

\usepackage{amsmath, amssymb, graphics, epsfig, graphicx, color}
\usepackage{epsf}
\usepackage{epstopdf}
\usepackage{subfig}

\newcommand{\nc}{\newcommand}
\nc{\ba}{\begin{eqnarray}}
\nc{\ea}{\end{eqnarray}}

\newcommand{\calR}{{\cal{R}}}

\newcommand{\calP}{{\cal{P}}}

\def\bfk{{\bf k}}

\def\bfq{{\bf q}}

\newcommand{\bfx}{{\bf{x}}}
\nc{\im}{{ \mathrm{Im} } }

\newcommand\be{\begin{equation}}
\newcommand\ee{\end{equation}}

\nc{\x}{{\bf{x}}}

\nc{\ep}{{\epsilon}}

\nc{\pp}{{\bf{p}}}

\nc{\nn}{\nonumber}
\nc {\by}{{\bf y}}
\nc{\vp}{\varphi}

\def\bk{{\bf k}}

\def\bq{{\bf q}}
\def\bx{{\bf x}}

\def \bfk{{\bk}}
\def \bfq{{\bq}
\def \bfx{{\bx}}}

\def \la {\langle}
\def \ra {\rangle}

\def \bfkb {\bfk_{\bot}}
\def \bfqb {\bfq_{\bot}}

\begin{document}

%%%%%%%%%%%%%%%%%%%%%%%%%%%%%%%%%%%%%%%%%%%%%%%%%%%%%
%\begin{flushright} {\footnotesize IC/2007/001}  \end{flushright}
\vspace{5mm}
\vspace{0.5cm}
\begin{center}

{\large \bf  Primordial anisotropies from cosmic strings during inflation 
}
\\[0.5cm]

{Sadra Jazayeri$^{1}$, 
Alireza Vafaei Sadr $^{2, 3}$,  Hassan Firouzjahi$^{1}$   }
 \vspace{0.3cm}

{\small \textit{$^1$School of Astronomy, Institute for Research in Fundamental Sciences (IPM) \\ P.~O.~Box 19395-5531, Tehran, Iran\\
}}
{\small \textit{$^2$ Department of Physics, Shahid Beheshti University, G.C., Evin, Tehran 19839, Iran
}}\\
{\small \textit{$^3$
D{\'e}partement de Physique Th{\'e}orique and Center for Astroparticle Physics, Universit{\'e} 
de Gen\`eve, 24 Quai Ernest Ansermet, 1211 Gen\`eve 4, Switzerland}}

\end{center}

\vspace{.8cm}

\hrule \vspace{0.3cm}
%{\small  \noindent \textbf{Abstract} \\[0.3cm]
%\noindent

%%%%%%%%%%%%%%%%%%%%%%%%%%%%%%%%%%%%%%%%%%%%%%%%%%%%%
\begin{abstract}

In this work we study the imprints of a primordial cosmic string on inflationary power spectrum. Cosmic string induces two distinct contributions on curvature perturbations power spectrum.  The first type of  correction
respects the  translation invariance while violating isotropy. This generates quadrupolar statistical anisotropy in CMB maps which is constrained by the Planck data. The second contribution breaks both homogeneity and isotropy, generating a dipolar power asymmetry in variance of temperature fluctuations with its amplitude falling on small scales. We show that the strongest  constraint on the tension of string is obtained from  the quadrupolar anisotropy and argue that the mass scale of underlying theory responsible for the formation of string can not be much higher than the GUT scale. The predictions of string for the diagonal and off-diagonal components of CMB angular power spectrum are presented.

\end{abstract}
%\vspace{0.5cm}  \hrule
%%%%%%%%%%%%%%%%%%%%%%%%%%%%%%%%%%%%%%%%%%%%%%%%%%%%%

\newpage
%%%%%%%%%%%%%%%%%%%%%%%%%%%%%%%%%%%%%%%%%%%%%%%%%%%%%
\section{Introduction}

The precise measurements of anisotropies on cosmic microwave background (CMB) temperature fluctuations and its polarization maps  \cite{Ade:2015lrj, Ade:2015xua, Planck:2013jfk} have provided strong supports for inflation as the leading theory  for early universe and generating the initial perturbations. The basic predictions of inflation that the CMB  perturbations to be nearly scale invariant, nearly adiabatic and nearly Gaussian are well consistent with these observations.

There are indications of anomalies on CMB maps as reported in Planck results \cite{Ade:2015lrj, Planck:2013jfk}  and also in earlier observations, such as the dipole asymmetry and the power suppressions on large scales. There are two different views as how to interpret these anomalies. One attitude is that these anomalies are not statistically significant and may be due to lack of precise data, 
unknown systematics or even methods of data analysis. This is mainly motivated from the fact that these anomalies are observed on low $\ell$ regions of CMB maps in which the effects of cosmic variance are non-negligible. It is possible that these anomalies are artifacts of poor statistics on large scales. In this view, no single anomaly is significant enough  to challenge the simple concordance model of early universe. It is argued that if a theoretical  model can address more than one anomalies at the same time, then these anomalies and the theory behind their generations become significant.
The other attitude is that these anomalies may be genuine and may hint to non-trivial inflationary dynamics. If so, understanding these anomalies may open new window to physics of primordial universe.  
This is particularly important if the anomalies persist in the current and future observations. In addition, if some theoretical models can address not only anomalies in CMB temperature maps but also provide independent predictions for CMB polarization maps and primordial tensor perturbations then it worth  studying these scenarios. 

In particular, the Planck data indicate the existence of hemispherical power asymmetry in the CMB maps \cite{Ade:2013nlj, Ade:2015hxq} which was observed earlier in the WMAP data too \cite{Eriksen:2003db, Eriksen:2006xr, Eriksen:2007pc}.  Fitting the temperature anisotropy  with a dipole modulation \cite{Gordon:2005ai} in the form 
\ba
\label{dipole}
\Delta T(\hat {\bf n}) = \overline{\Delta T(\hat {\bf n})} \left( 1 + A_d \hat {\bf n} \cdot \hat {\bf p} \right)  \, ,
\ea
the Planck data  found the dipole  amplitude $A_d \simeq 0.06$ with the preferred direction $\hat {\bf p} $
towards the southern hemisphere with respect to the galactic plane. One interesting feature of dipole asymmetry is that the amplitude of dipole shows strong scale-dependent such that it falls off rapidly on smaller scales, say for $\ell \ge 100$.   The effects of dipole asymmetry in CMB data and large scale structure have been further investigated in \cite{Akrami:2014eta, Aiola:2015rqa, Mukherjee:2015mma, Mukherjee:2015wra, Adhikari:2014mua, Quartin:2014yaa, Shiraishi:2016wec, Shiraishi:2016omb, Baghram:2014nha, Hassani:2015zat, Zibin:2015ccn, Yasini:2016dnd}.

With these discussions in  mind  there have been significant interests to address the nature of dipole asymmetry  in recent years. One interesting proposal for the mechanism of dipolar asymmetry  is the idea of long mode modulations  \cite{aniso-longmode}. In this picture it is assumed that there exists a mode $k_L$ which is much longer than the Hubble radius during inflation. This long mode generates the power asymmetry by modulating the background inflationary parameters such as the inflaton field or its velocity or by modulating the surface of end of inflation. Unfortunately  this proposal does not work  in simple models of inflation such as in single field scenarios. Based on the single field non-Gaussianity consistency condition \cite{Maldacena:2002vr},  it is  shown in \cite{Namjoo:2013fka} that the amplitude of dipole modulation is controlled by  the amplitude of local-type non-Gaussianity $f_{NL}$. Consequently,  in single field  inflationary models  with small (actually zero) $f_{NL}$,  dipole asymmetry with large enough amplitudes can not be generated. For this idea to work, one has to consider models  beyond simple slow roll scenarios  such as  curvaton model, iso-curvature perturbations etc.  For a list of various theoretical works based on long mode modulation and related ideas to generate dipole asymmetry see \cite{various}.

Alternatively, the idea of using primordial defects during inflation to generate power asymmetry has been employed in \cite{Jazayeri:2014nya} and \cite{Firouzjahi:2016fxf}. In \cite{Jazayeri:2014nya} it is assumed that there exists a domain wall during inflation which causes the asymmetry. It is shown that large dipole with non-trivial scale dependent can be generated while the amplitude of higher multipoles are suppressed as required from the Planck data.  This idea was extended in  \cite{Firouzjahi:2016fxf} to the case of a primordial massive defect, such as a monopole or 
black hole,  during inflation  to generate power asymmetry. The presence of a massive defect breaks the 
translational invariance maximally while keeping the rotation invariance intact. The structure of power asymmetry is somewhat non-trivial as one also generates inhomogeneities  in primordial power spectra.

Another anomaly which captured significant interests in recent year is quadrupolar statistical anisotropy.
Unlike the hemispherical (dipolar) asymmetry defined in Eq. (\ref{dipole}), the quadrupolar statistical anisotropy represents  anisotropy at individual points. Specifically, if one divides the CMB sphere in two opposite hemispheres then both hemisphere are statistically the same while points on the same or opposite hemisphere can have different power.   The quadrupolar statistical anisotropy in curvature perturbation power spectrum $P_\calR  $ is usually parametrized in Fourier space $\bfk$ via \cite{Ackerman:2007nb, Pullen:2007tu}
\ba
\label{g-star}
P_\calR (\bfk) =  P^{(0)}_\calR (\bfk) \big( 1+ g_* ( {\bf \hat m} \cdot \bf \hat k)^2 \big )  \, ,
\ea
in which $P^{(0)}_\calR (\bfk) $ is the dominant isotropic power spectrum,  ${\bf \hat m}$ represents the  preferred (anisotropic) direction in the sky and  $g_*$
is the amplitude of quadrupolar anisotropy. Constraints from Planck data \cite{Kim:2013gka, Ade:2015lrj, Planck:2013jfk} implies  $| g_*| \lesssim 10^{-2}$.  

The best known mechanism to generate quadrupolar statistical anisotropy is the scenario  of anisotropic inflation based on the dynamics of a $U(1)$ gauge field during inflation,  see for example  \cite{Watanabe:2009ct, Watanabe:2010fh, Bartolo:2012sd, Shiraishi:2013vja, Emami:2015qjl,  Abolhasani:2013zya, Rostami:2017wiy}. 
In this mechanism a background electric field is turned on during inflation so the background geometry is in the form of Bianchi I metric. If one couples the gauge field with the inflaton field appropriately, then one can reach the attractor regime in which the electric field energy density reaches a sub-leading but a constant fraction of the total energy density. This can leads to a small amount of quadrupole anisotropy.

Mathematically speaking, the quadrupolar statistical anisotropy given in Eq. (\ref{g-star}) is defined
in Fourier space while the hemispherical asymmetry in Eq. (\ref{dipole}) is defined in real space. In order to prevent confusion, we refer to former as the statistical anisotropy while the latter is called power asymmetry or dipolar asymmetry.

In this work, we extend the motivation of \cite{Jazayeri:2014nya, Firouzjahi:2016fxf} to the case of a cosmic strings during inflation. Our goal is to calculate the corrections in curvature power spectrum and 
to look for the amplitude, shape and scale dependence of the induced anisotropy and asymmetry. The  imprints of primordial defects during inflation for various motivations 
have been studied in \cite{Carroll:2008br, Tseng:2009xw, Prokopec:2010nm, Wang:2011pb, Cho:2009en, Cho:2014nka}. In particular in  \cite{Tseng:2009xw} the correction to curvature perturbation power spectrum induced by  a  cosmic string during inflation is obtained. In this work we build and extend on the results of \cite{Tseng:2009xw}  to obtain the imprints of a cosmic string during inflation. In the presence of a cosmic string, both of the translation and rotation invariances are partially broken  while a subset of these symmetries are left intact. More specifically,  the translation and the rotation along  the string are still invariant. Consequently, we expect that  corrections in power spectrum to retain the translation  and the rotation invariances only along the string.  As a result, as we shall see,  cosmic string  has the unique property to  generate both quadrupole anisotropy and power  asymmetry though with complicated shapes.

Before closing this section, we comment that independent of the observational significance of anisotropy and asymmetry,  the idea of looking for the imprints of defects during inflation is well-motivated. Indeed, the formation of defect is a generic feature of symmetry breaking which are expected to happen at various scales on the history of early universe \cite{Kibble:1976sj}.  In particular, the idea of strings in early universe is interesting. In models of inflation constructed from  string theory, such as in brane inflation, cosmic strings 
are copiously generated  at the end  or during inflation when a pair of brane and anti-brane annihilate each other \cite{Sarangi:2002yt, Jones:2003da, Copeland:2003bj, Firouzjahi:2005dh, Firouzjahi:2006vp, Chernoff:2014cba}. 
These are either fundamental strings (F-strings) or D1-branes (D-strings) which have different charges and couplings. They can combine to form junctions of $(p, q)$ strings which can have non-trivial implications for lensing and evolution of the networks of cosmic superstrings, for a review see \cite{HenryTye:2006uv}.

%%%%%%%%%%%%%%%%%%%%%%%%%%%%%%%%%%%%%%%%%%%%%%%%%%%%
\section{Curvature perturbations power spectrum }
\label{setup}

In this section we present our setup of a cosmic string during inflation. The motivations and the logics are
similar to \cite{Jazayeri:2014nya} and \cite{Firouzjahi:2016fxf}. It is assumed that inflation is driven by a scalar field, the inflaton field $\phi$, which is slowly rolling on its nearly flat potential $V(\phi)$. Therefore, the dominant source of energy density is given by the potential $V$. The string is assumed to be a sub-dominant source of energy and its effects can be treated perturbatively compared to those of inflaton field. We consider the idealized situation where the string's length is much larger than the Hubble radius during inflation so for practical purposes it is treated as a string with  infinite length.  We assume that all wiggles along the length of strings  are wiped out so it can be parameterized by its tension $\mu$.
As usual, the relevant dimensionless parameter in the studies of cosmic string is the parameter $G \mu$
which measures the tension of string in units of Newton constant $G$. In order for our perturbative approach to be consistent and the contribution of string to energy density in a given Hubble radius $H^{-1}$ to be sub-dominant compared to the inflaton potential, we require that $( \mu H^{-1}) H^3 \ll V$ which is equivalent to $G \mu \ll 1$. In addition, we work in the limit that the thickness of string is negligible so it can be treated as a line of distribution of energy with the tension $\mu$. 

If we consider the physical assumption that the string is formed because of a $U(1)$ symmetry breaking during inflation, then the thickness of string is related to the energy scale of symmetry breaking which is at the order  $1/\sqrt \mu$. Therefore, assuming the thickness of string to be much smaller than the Hubble radius during inflation, we require $1/\sqrt \mu \ll H^{-1}$ which in turn translates into 
$ G \mu \gg (H/M_P)^2$ in which $M_P= 1/8 \pi G$ is the reduced Planck mass.
Combining both conditions, we require $  (H/M_P)^2 \ll G \mu \ll 1$. For typical models of inflation we expect $H/M_P \lesssim 10^{-5}$ so the above condition can be easily satisfied for $G \mu \lesssim 10^{-2}$. 

The upper bound on the tension of string is  $G\mu \lesssim 10^{-7}$ if a network of cosmic strings is assumed to generate parts of temperature anisotropies in CMB maps, for some recent works on this direction see for example  \cite{Lizarraga:2016onn, Charnock:2016nzm, Lizarraga:2014xza, Lazanu:2014eya}.  However, this bound does not apply to our case, since we do not consider a network of strings to generate perturbations on CMB after inflation. In our picture, we have one string in  a Hubble horizon during inflation. In addition, in order not to complicate the thermal history of universe after inflation, we assume that the string is decayed to relativistic particles during reheating so all its energy goes to radiation after inflation. In general the latter assumption may not be necessary so it may be relaxed if one is interested in the presence of string  after inflation. 

We are interested in corrections to curvature perturbation power spectrum induced from string. Following the logics of \cite{Jazayeri:2014nya, Firouzjahi:2016fxf} the dominant contribution in comoving curvature perturbations $\calR$ is given by inflaton field via 
\ba
\label{calR-def}
\calR=- \frac{H}{\dot{\phi}}\delta \phi  \, ,
\ea
in which $\delta \phi$ is the quantum fluctuations associated with the inflaton field.  
There are two  types of  contributions from string which in principle one has to take into account. First, the definition of curvature perturbation $\calR$ in the presence of string is modified so there will be additional term in $\calR$ beyond the leading term given in Eq. (\ref{calR-def}). Second, the string modifies the background geometry. As is well-known the geometry around a straight string is locally flat while it modifies the geometry  globally causing a deficit angle at the order $G \mu$ around string \cite{Vilenkin:2000jqa}. As argued in  \cite{Jazayeri:2014nya, Firouzjahi:2016fxf} the first contribution in curvature power spectrum are 
at the order $G\mu \sqrt{\epsilon_H}$ in which $\epsilon_H$ is the slow-roll parameter $\epsilon_H \equiv -\dot H/H^2$. Intuitively speaking, the first contribution comes from the gravitational back-reactions of string on inflaton dynamics which necessarily has both of small parameters $G \mu$ and $\sqrt \epsilon_H$.  However, the second contribution  is the direct contribution of string into background geometry which is  at the order $G\mu$ as we calculate below. Therefore,  in the slow-roll limit where $\epsilon_H \ll1$, the leading correction is from  the second contribution, i.e. the direct contribution of string in geometry.  This in turn induces a correction in Hamiltonian and its effects on curvature perturbation power spectrum  can be calculated using the perturbative in-in formalism \cite{Maldacena:2002vr, Weinberg:2005vy}. 

With these discussions in mind, now we proceed to study the effects of cosmic strings on background geometry. As mentioned before, in flat background the geometry around string is locally flat while a deficit  is induced around string. In an inflationary background with a near dS background, one expects the above picture to hold and the string only to induce a deficit angle without changing the local geometry. Specifically, assuming the infinite string is extended along the $z$ direction, 
the vacuum solution of string in dS background in polar coordinate $(\rho, \phi, z)$
have been obtained to be \cite{Abbassi:2003fh}
\be
ds^2=-dt^2+a(t)^2\Big(d\rho^2+(1-4G\mu)^2 \rho^2d\phi^2+dz^2 \Big) \, ,
\ee
in which $a(t) = \exp( H t)$ is the scale factor in the dS background. To leading order in slow-roll correction we have neglected the variation of $H$ which results in sub-leading corrections to our analysis, i.e. at the order $G\mu \sqrt{\epsilon_{H}}$ or higher. Note that the metric above is written non-perturbatively to all orders in $G \mu$. However, we are only interested in corrections to leading order in $G\mu$ so we shall expand the above metric to first order in $G \mu$. Also note that the above metric satisfies the intuition  that the string does not change the local metic of spacetime and only induces a deficit angle equal to $8 \pi G \mu$. 

It is more convenient to work with the Cartesian coordinate system in which the above metric is transformed into  
\be
\label{metric}
ds^2=-dt^2+a(t)^2 \Big(d\bfx^2-\frac{\epsilon}{\rho^2}(x^2dy^2+y^2dx^2-2x\, y \, dx\, dy) \Big )
\ee
in which $\rho^2 = x^2 + y^2$ and  we have defined the  small dimensionless parameter $\epsilon$ via 
\be
\epsilon=8G\mu  \, .
\ee

We need to calculate the interaction Hamiltonian. For this purpose, we write down the action of inflaton field in the presence of cosmic string. In our treatment the inflaton field  feels the presence of string via the deformation of background geometry induced by cosmic string as given in Eq. (\ref{metric}). Note that in the limit where we neglect the gravitational back-reactions of string on inflaton field,  we can treat the scalar field as a nearly massless scalar field with the amplitude of quantum fluctuations $H/2\pi$. The  rollings of inflaton and its mass induces corrections at the order $\epsilon \sqrt \epsilon_H$ in anisotropic power spectrum  which we neglect as  mentioned before.  

The action of  a (nearly) massless inflaton field in the presence of cosmic string encoded in  the geometry  (\ref{metric}) is given by 
\ba
{S}=- \frac{1}{2} \int d^4x \sqrt{-g}\,  g^{\mu\nu} \partial_{\mu}\delta \phi\partial_{\nu}\delta \phi \, .
\ea
Calculating the inverse metric $g^{\mu \nu}$ and the determinant $\sqrt{-g}$ to leading order in $\epsilon$, we have  
\ba
\sqrt{-g}=a^3(1-\frac{\epsilon}{2})+{\cal O}(\epsilon^2) \,  ,
\ea
and  
\ba
\delta g^{xx}=\epsilon \frac{y^2}{a^2\rho^2}\, , \qquad
\delta g^{yy}= \epsilon \frac{x^2}{a^2 \rho^2} \, , \qquad 
\delta g^{xy}=\delta g^{yx}=-\epsilon \frac{xy}{a^2 \rho^2} \, .
\ea

Since the interaction terms in the Lagrangian contain solely space derivatives, the Hamiltonian density ${\cal H}_I$ simply equals the Lagrangian density $-{\cal L }_I$. Plugging back  the above results into the action,  the leading order interaction Hamiltonian is obtained  to be 
\ba
\label{HI}
H_I =\frac{\epsilon a(t)}{2} \int d^3\bfx \frac{ \left(x\partial_y\delta \phi-y\partial_x\delta \phi \right)^2}{(x^2+y^2)}  \, .
\ea

Because we are interested in curvature perturbation power spectrum in Fourier space, we calculate $H_I$ in the Fourier space, yielding  
\ba
H_I=- \frac{a(t) \epsilon}{2(2\pi)^6} \int d^3\bfx \,  d^3\bfk \,  d^3\bfq \,  \frac{\delta  \phi_k \delta \phi_q}{x^2+y^2}\, ( y k_x-xk_y) \,  (yq_x-xq_y)  e^{ i(\bfk+\bfq).\bfx  } \, ,
\ea
in which $\delta \phi_k$ is the amplitude of $\delta \phi $ fluctuations in Fourier space.   From the above expressions for $H_I$ we see that the system enjoys the remnant  translation and rotation  symmetries
around  the $z$ direction, the orientation of string.

Now using the standard in-in formalism \cite{Maldacena:2002vr, Weinberg:2005vy}, the corrections in two-point correlations of inflaton field induced by cosmic string to leading order in $\epsilon$ is  obtained to be   
\ba
\label{Dyson}
\Delta \big \langle \delta \phi_\bfk (t_e) \delta \phi_\bfq(t_e)\big \rangle &=& i \int_{0}^{t_e} dt' \big \langle \left[ \, H_I(t'),  \, \delta \phi_\bfk \delta \phi_\bfq  \, \right ] \big \rangle  \nonumber\\
&=&- 2 \im \int_0^{t_e} d t' \big \langle   H_I(t') \delta \phi_{\bk} (t_e) \delta \phi_{\bq}(t_e) 
\big \rangle  \, ,
\ea
in which $t_e$ indicates the time of end of inflation.

Going to conformal time $d \eta = dt/a(t)$ we obtain
\ba
\label{Delta-1}
\Delta  \langle \delta \phi_\bfk (t_e) \delta \phi_\bfq(t_e) \rangle &=&
2 \epsilon ( 2 \pi) \delta (k_z + q_z)  \nonumber\\
&\times&  \int d \eta' a^2(\eta')  h(\bfk_\perp, \bfq_\perp)   \, 
\im \Big(  \delta \phi_k (\eta') \delta \phi_q(\eta') \, \delta \phi_k^*(\eta_e) \,  \delta \phi_q^*(\eta_e)
\Big)  \, ,
\ea
in which $\bfk_\perp$ represents the projection of $\bfk$ on the $xy$ plane which is perpendicular to the orientation of string and  we have defined the  function $h(\bfk_\perp, \bfq_\perp) $ via
\ba
h(\bfk_\perp, \bfq_\perp)   \equiv  \int d^2 \bfx \,  \frac{\exp \big(i(\bfk+\bfq)_\perp.\bfx_\perp \big)}{ x^2 + y^2}  
\Big[ x^2 q_y k_y + y^2 q_x k_x - x y ( q_x k_y + q_y k_x) 
\Big] \, .
\ea
One can easily check that $h(\bfk_\perp, \bfq_\perp) = h(\bfq_\perp, \bfk_\perp)$.  Also note that the 
delta function $\delta (k_z + q_z) $ in Eq. (\ref{Delta-1}) is a manifestation of translation invariance along the string. 

Using the following form for the wave function of inflaton field 
\ba
\nonumber
\delta{\phi}_k=\frac{H}{\sqrt{2k^3}}(1-ik\eta)\exp(ik\eta) \, ,
\ea
the integrand  in Eq. (\ref{Delta-1}) simplifies to 
\ba
2 \epsilon \int_{-\infty}^{0} \frac{d \eta'}{H^2 \eta'^2} \frac{H^2}{ \sqrt{k^3 q^3}}    \im \Big[ ( 1- i k \eta') ( 1- i q \eta') e^{i (k+ q) \eta'}   \Big]  \, .
\ea 
Now using the relations
\ba
\im \int_{-\infty {(1-i\epsilon)}}^{0} d \eta'   e^{i (k+ q) \eta'} = \frac{-1}{ k+ q}
\ea
and
\ba
\im \int_{-\infty {(1-i\epsilon)}}^{0} d \eta' \big( \frac{1}{\eta'^2} - \frac{i (k+ q)}{\eta'} \big)   e^{i (k+ q) \eta'} =  - (k+ q) \, ,
\ea
the integral  over $\eta'$ in Eq. (\ref{Delta-1}) is calculated yielding 
\ba
\label{Delta-2}
\Delta \big \langle \delta \phi_\bfk (t_e) \delta \phi_\bfq(t_e)\big \rangle &=& 
\frac{ \epsilon \pi H^2}{k^3 q^3}  \delta (k_z + q_z) \Big(  \frac{k q}{k+q} - (k+ q)  \Big)  \nonumber\\
~~~~~~\qquad \qquad && \times \Big[ \sum_{i \neq j}  k_jq_j F_{ii}{(\bk_\bot+\bq_\bot)}-k_iq_j F_{ij}{(\bk_\bot+\bq_\bot)} \Big]  \, ,
\ea
in which  for $i, j=1, 2$ we have defined 
\be
F_{ij}(\bfk_\perp)\equiv\int d^2\bfx \exp(i\bfk_\perp.\bfx_\perp)\frac{x_ix_j}{x^2+y^2}  \, ,  \quad \quad i,j=1,2 \, .
\ee
With some efforts one can check that  \cite{Tseng:2009xw}
\be
\label{Fii}
F_{ij}(\bfk_\perp)=2\pi^2\delta_{ij}\delta^2(\bfk_\perp)+\frac{4\pi}{k_1^2+k_2^2}\left( \frac{\delta_{ij}}{2}-\frac{k_ik_j}{k_1^2+k_2^2} \right)  \, .
\ee

Plugging the above form of $F_{ij}(\bfk_\perp)$ in Eq. (\ref{Delta-2}), and noting that the curvature perturbation $\calR$ is related to $\delta \phi$ via $\calR = -H \delta \phi/\dot \phi$,  the corrections of cosmic strings in curvature perturbations two point correlation function is  obtained to be   
\ba
\label{Delta-3}
\Delta \big \langle \calR_\bfk (t_e) \calR_\bfq(t_e)\big \rangle &=& 
- \epsilon \pi  \Big( \frac{H^2}{\dot \phi} \Big)^2
\Big( \frac{k^2 + q^2 + kq}{  k^3 q^3 (k+ q)}  \Big)  \delta (k_z + q_z)
 \Big[ 2 \pi^2 \bfk_\perp \cdot \bfq_\perp \delta^2 ( \bfk_\perp +  \bfq_\perp) \nonumber\\
&&~~~~~~~~~~~~~~~~~~~ + \frac{2 \pi \bfk_\perp \cdot \bfq_\perp  }{( \bfk_\perp +  \bfq_\perp)^2 } +
 \frac{4 \pi  }{( \bfk_\perp +  \bfq_\perp)^4 }   ( k_x q_y- k_y q_x)^2
\Big] \, .
\ea

This is the main result of this section.  The structure of the symmetries of two point correlation function is somewhat non-trivial.  The full $SO(3)$ rotation is broken to the subset of two-dimensional rotation in the $xy$ plane 
as one can easily see that all three terms above are invariant under rotation only around the string. As for translation invariance, only the first term above retains the full three-dimensional translation invariance because it has the three dimensional Dirac delta function $\delta ^3(\bfk + \bfq)$.
The last two terms in the big bracket breaks the translation invariance in the plane perpendicular to string as they have only $ \delta (k_z + q_z)$. Since the string loses the full rotation and translation invariances, its corrections in curvature perturbation power spectrum is a mixture of anisotropies and inhomogeneities. 
Therefore, the asymmetries generated by cosmic string is more complicated than the simple dipole asymmetry modeled by Eq. (\ref{dipole}) and can not be  captured just by the dipole amplitude $A_d$.

One can check that our result for two point correlation in Eq. (\ref{Delta-3}) agrees with the result obtained in \cite{Tseng:2009xw}. Indeed, manipulating the terms inside the big bracket in Eq. (\ref{Delta-3}) one can show that
\ba
\label{Delta-Wise}
\Delta \big \langle \calR_\bfk (t_e) \calR_\bfq(t_e)\big \rangle &=&
- \epsilon \pi  \Big( \frac{H^2}{\dot \phi} \Big)^2  \Big( \frac{k^2 + q^2 + kq}{  k^3 q^3 (k+ q)}  \Big)  \delta (k_z + q_z)
 \Bigg[ 2 \pi^2 \bfk_\perp \cdot \bfq_\perp \delta^2 ( \bfk_\perp +  \bfq_\perp) \nonumber\\
&& - \frac{4 \pi   }{( \bfk_\perp +  \bfq_\perp)^2 } 
\Big(  \frac{ \bfk_\perp \cdot \bfq_\perp}{2} - \frac{\bfk_\perp \cdot ( \bfk_\perp +  \bfq_\perp)\,  \bfq_\perp \cdot ( \bfk_\perp +  \bfq_\perp)  }{( \bfk_\perp +  \bfq_\perp)^2}
\Big) \Bigg]  \, ,
\ea  
as obtained in \cite{Tseng:2009xw}.

Now adding the leading isotropic and homogenous contribution from the inflaton field itself, the total two point correlation function is given by 
\ba
\label{power-total}
 \big \langle \calR_\bfk (t_e) \calR_\bfq(t_e)\big \rangle =
  \Big( \frac{H^2}{\dot \phi} \Big)^2 
\Bigg[ \frac{(2 \pi)^3}{k^3} \delta^3 ( \bfk + \bfq)   - \epsilon \pi   \Big( \frac{k^2 + q^2 + kq}{  k^3 q^3 (k+ q)}  \Big)  \delta (k_z + q_z) \times~~~~~~~~~~~~~
\nonumber\\
 \times \Big[ 2 \pi^2 \bfk_\perp \cdot \bfq_\perp \delta^2 ( \bfk_\perp +  \bfq_\perp) 
+ \frac{2 \pi \bfk_\perp \cdot \bfq_\perp  }{( \bfk_\perp +  \bfq_\perp)^2 } +
 {\frac{4 \pi  }{( \bfk_\perp +  \bfq_\perp)^4 }   ( k_x q_y- k_y q_x)^2
\Big] \Bigg] \, }.
\ea
%Note that, the contribution from inflaton field is invariant under the full three dimensional rotation and translation as expected. 

Having obtained the corrections from cosmic string in two point functions in Fourier space as in Eq. (\ref{Delta-3}) or Eq. (\ref{Delta-Wise}) we can use them to look for the predictions of cosmic string on the CMB temperature maps and obtain some estimations of the preferred values of the model parameters. 

%We will do it in section \ref{numeric}.  However, before then  we calculate the variance of curvature perturbations which provides insights for the form of anisotropies and asymmetries generated by cosmig string.   

%%%%%%%%%%%%%%%%%%%%%%%%%%%%%%%%%%%%%%%%%%%%%%%%%%%%
\section{Quadrupole anisotropy}
\label{quadrupole}

As we discussed above, the corrections in power spectrum induced from cosmic string  have two distinct contributions. The first term in Eq. (\ref{Delta-3}) retains the three-dimensional translation invariance while the last two terms in Eq. (\ref{Delta-3}) are translation invariant only along the string. 

Interestingly we see that the first term in Eq. (\ref{Delta-3}) has the structure of a quadrupolar anisotropy as introduced  in Eq. (\ref{g-star}) in which the anisotropic (preferred) direction is the orientation of cosmic string.
As can be seen, the contribution of quadrupolar anisotropy is quite different than the contribution of 
last two terms  in Eq. (\ref{Delta-3}) which mostly mimic a dipolar asymmetry. As we discussed before, the quadrupolar statistical anisotropy is  associated with anisotropy at each point on CMB map while each CMB hemisphere has statistically the same power as the opposite hemisphere. 

To calculate the amplitude of quadrupolar anisotropy $g_{*}$ we compare the quadrupole term in power spectrum 
with the   isotropic power spectrum $P^{(0)}_\calR $ given by the first term in Eq. (\ref{power-total}), obtaining 
\ba
\label{g-star-val}
g_* = -\frac{3 \epsilon}{8} \, .
\ea
The minus sign above is from the fact  that $\sin^2 \theta = 1- \cos^2 \theta$. 

This is an interesting prediction. We see that cosmic string induces a quadrupole anisotropy in CMB maps which can be tested directly by cosmological observations. In particular, constraints 
from Planck observations \cite{Kim:2013gka, Ade:2015lrj, Planck:2013jfk}  implies   $| g_*| \lesssim 10^{-2}$, yielding  $\epsilon \lesssim 10^{-2}$. Therefore, the scale of symmetry breaking responsible for the formation of string during inflation can not be much higher than the GUT scale. For example, if we assume that the cosmic string in early universe are in the forms of  D- or F-string from string theory, then the mass scale of string theory can not be much higher than the GUT scale.   In addition, we  see that the sign of $g_*$ is negative in our setup. Curiously, the sign of $g_*$ is also negative in all known models of anisotropic inflation \cite{Rostami:2017wiy}.

%%%%%%%%%%%%%%%%%%%%%%%%%%%%%%%%%%%%%%%%%%%%%%%%%%%%
\section{Variance of curvature perturbations}
\label{variance}

Since the last two terms in Eq. (\ref{Delta-Wise}) are not fully homogeneous, we expect them to induce an effective power asymmetry in CMB maps. The structure of these terms are too complicated to be used directly
in an analytical study. A practically useful analytical tool is to look for the variance   of curvature perturbations in real space  $\langle \calR(\bfx)^2 \rangle$.   This provides insights about the magnitude and the form of power asymmetry  generated by cosmic string in temperature fluctuations. Following the analysis of \cite{Akrami:2014eta}, Planck team has used the variance of temperature fluctuations (which is linearly related to $ \calR(\bfx)$) as one of the measure of dipole asymmetry \cite{Ade:2015hxq}. With this motivation in mind we  calculate $\langle \calR(\bfx)^2 \rangle$ for our setup.  

The dominant contribution in variance comes from the inflation field which is given by the usual curvature perturbation power spectrum.  Denoting this dominant isotropic and homogeneous contribution by  
$\langle \calR(\bfx)^2 \rangle^{(0)}$, we have 
\ba
\label{var-iso}
\langle \calR(\bfx)^2 \rangle^{(0)} = \int d \ln k\, \left(  \frac{k^3}{2 \pi^2} | \calR(\bfk)|^2 \right)  = \int d \ln k \, {\calP_\calR}^{(0)} \, ,
\ea
where ${\calP_\calR}^{(0)}$ is the isotropic power spectrum. For single field slow roll model it is equal to ${\calP_\calR}^{(0)} = ( H^{2}/2\pi \dot \phi)^{2}$.   Taking the power spectrum to be nearly scale invariant the variance in  Eq. (\ref{var-iso}) takes the form $\langle \calR(\bfx)^2 \rangle^{(0)} ={\calP_\calR}^{(0)} 
\ln (k_{S}/k_{L}) $ in which $k_{L}$ and $k_{S}$ respectively are the IR and UV cutoffs of the system. Note that the logarithmic divergence is a feature of neglecting inflaton's  mass. Taking into account the effects of inflation's small mass, the logarithmic divergence is expected to disappear. 

The correction in variance induced by cosmic string is given by 
\ba
\Delta \langle \calR(\bfx)^2 \rangle = \frac{1}{(2 \pi)^6} \int\int\textrm{d}^3\bfk\textrm{d}^3\bfq \,  e^{i (\bfk+ \bfq)\cdot \bfx} \Delta \langle \calR_{\bfk} \calR_{\bfq} \rangle \, ,
\ea
in which  $\Delta \langle \calR_{\bfk} \calR_{\bfq} \rangle$ is calculated in  Eqs. (\ref{Delta-3}) or (\ref{Delta-Wise}). The first term in Eq. (\ref{Delta-Wise}) is fully  translation invariant so as expected 
it does not generate any position dependence; it only modifies the leading isotropic variance. Denoting the contribution from  the first term in Eq. (\ref{Delta-Wise}) to variance by $\Delta \langle \calR(\bfx)^2 \rangle_{{\rm hom.}}$ we obtain
\ba
\label{var-hom}
\Delta \langle \calR(\bfx)^2 \rangle_{{\rm hom.}} &=& -\frac{3 \epsilon}{16}   \left( \frac{H}{\dot \phi} \right)^2
\int_{-\infty}^{\infty} d k_z \int_0^\infty d k_\perp \frac{k_\perp^3}{( k_\perp^2 + k_z^2)^{5/2}} 
\nonumber\\
&=& - \epsilon \int d \ln k \, {\calP_\calR}^{(0)} \, .
\ea
As expected, this has the same structure as the leading homogeneous variance given in Eq. (\ref{var-iso}). 
The constraint from the quadrupole anisotropy $\epsilon \ll 1$ guarantees that the corrections in isotropic and homogeneous variance induced from the first term in (\ref{Delta-Wise}) is sub-leading compared to contribution from inflaton field.  The interesting feature is that this contribution from string has opposite sign compared to contribution from inflaton. This may be a good news to address the shortage of power on low $\ell$ as observed in Planck data. However, a careful data analysis must be performed to see whether the string can address the shortage of power on large scales while not changing the power on smaller scales and at the same time satisfying the constraints from the quadrupole statistical anisotropy. 

The remaining two terms in Eq. (\ref{Delta-Wise}) violates the translation invariance in $xy$ plane and contribute nontrivially in variance. Denoting these contributions by 
$\Delta \langle \calR(\bfx)^2 \rangle_{{\rm asym.}}$ we have 
\ba
\label{var-asym1}
\Delta \langle \calR(\bfx)^2 \rangle_{{\rm asym.}} &=& -\ep  \Big( \frac{H^{2}}{\dot \phi} \Big)^{2}
\int  \frac{d^2\bfk_{\bot}  d^2\bfq_{\bot}dk_z }{(2\pi)^6}  e^{i (\bfkb+\bfqb).\bfx_{\bot}} \frac{4\pi}{(\bfkb+\bfqb)^2}
\Big( \frac{k^2+q^2+kq}{k^3q^3(k+q)}\Big)  |_{q_z=-k_z} \nn \\ 
  &\times&  \Big[  \frac{1}{2}\bfk_{\bot}.\bfq_{\bot}-\frac{1}{(\bfk_{\bot}+\bfq_{\bot})^2} \bfk_{\bot}.(\bfk_{\bot}+\bfq_{\bot}) \bfq_{\bot}.(\bfk_{\bot}+\bfq_{\bot})\Big]   \, ,
\ea
in which the delta function $\delta (k_{z} + q_{z})$ have been  used to remove the integration over $q_z$. 
This also means that inside the integral we have $k= \sqrt{\bfk_\perp ^2 + k_z^2} $ and  $q= \sqrt{\bfq_\perp ^2 + k_z^2} $. 

Fortunately the integral over $k_z$ can be taken analytically where 
\ba
\int_{-\infty}^{\infty} \frac{\bfkb^2+\bfqb^2+2k_z^2+(\bfkb^2+k_z^2)^{1/2}(\bfqb^2+k_z^2)^{1/2}}{(\bfkb^2+k_z^2)^{3/2}(\bfqb^2+k_z^2)^{3/2}\Big[ (\bfkb^2+k_z^2)^{1/2}+(\bfqb^2+k_z^2)^{1/2}\Big ]}dk_z 
=\frac{2}{k_\bot ^2 q_\bot ^2} \, .
\ea
Plugging this in Eq. (\ref{var-asym1}), yields  
\ba
\label{var-asym2}
\Delta \langle \calR(\bfx)^2 \rangle_{{\rm asym.}}& =& -8 \pi \epsilon \Big( \frac{H^{2}}{\dot \phi} \Big)^{2} \int \frac{d^2\bfk_{\bot}d^2\bfq_{\bot}}{(2\pi)^6} \frac{ e^{ i (\bfkb+\bfqb).\bfx_{\bot} }}{(\bfkb+\bfqb)^2 k_\bot^2 q_\bot^2}\\ \nn 
~~~~~~~~~~~~~~~~~~~~~~~~~~~~~&\times& \Big [  \frac{1}{2}\bfk_{\bot}.\bfq_{\bot}-\frac{1}{(\bfk_{\bot}+\bfq_{\bot})^2} \bfk_{\bot}.(\bfk_{\bot}+\bfq_{\bot}) \bfq_{\bot}.(\bfk_{\bot}+\bfq_{\bot})\Big ]  \, .
\ea

%Below we calculate the above integral analytically. 
The above expression for variance looks too complicated to be handled analytically.  
However, useful information can be obtained by looking  at its asymptotic behaviors. It is easy to see that the integral above 
has no UV divergence as the integrand oscillate rapidly yielding a finite UV contributions. As for the IR behavior, we note that the integral is independent of scale in the sense that if we rescale all momenta and the measures by the factor $|\bfx_{\bot}|$, then the integral and the measure remain independent of scale while all scale dependents appear at the lower cutoff of the integral, i.e. it appears at the IR cutoff of the integral. Using this insight, we rescale all momenta by $|\bfx_{\bot}|$ writing the asymmetric variance as
\ba
\label{var-asym3}
\Delta \langle \calR(\bfx)^2 \rangle_{{\rm asym.}}& =& -8 \pi \epsilon \left( \frac{H^{2}}{\dot \phi} \right)^{2} 
\int_{|\bfkb,\bfqb|>\frac{\rho}{L}}  \frac{d^2\bfk_{\bot}d^2\bfq_{\bot}}{ (2\pi)^6 k_\bot^2 q_\bot^2} e^{ i (\bfkb+\bfqb).\hat{\bfx}_{\bot}} \frac{1}{(\bfkb+\bfqb)^2 }\nonumber\\
~~~~~~~~~~~~~~~~~~~~~~~~~~~~~&\times& \Big [  \frac{1}{2} \bfk_{\bot}.\bfq_{\bot}-\frac{1}{(\bfk_{\bot}+\bfq_{\bot})^2} \bfk_{\bot}.(\bfk_{\bot}+\bfq_{\bot}) \bfq_{\bot}.(\bfk_{\bot}+\bfq_{\bot})\Big ]  \, .
\ea
Here $\rho \equiv x_\perp = \sqrt{x^2 + y^2}$ is the perpendicular distance from the point $\bfx$ on the CMB sphere to the string and  $L$ represents the IR comoving cutoff of the setup, the size of an imaginary  box  which is bigger than  the observable Universe.  

The scaling of the integrand above suggests that the dominant contribution in Eq. (\ref{var-asym3})  comes from the IR region in which $k, q$ approaches their IR lower end. Using this insight, we can easily obtain the order of magnitude of the above integral. For example, for the first term in big bracket above we have
\ba
\label{asym1}
 \int \frac{d^2\bfk_{\bot}d^2\bfq_{\bot}}{ k_\bot^2 q_\bot^2}  \frac{\bfk_{\bot}.\bfq_{\bot}}{(\bfkb+\bfqb)^2 }
 &=& \int   d\phi_1  d \phi_2  \, d k_\perp \,  d q_\perp     \frac{  \cos (\phi_1 -\phi_2) }{ k_\perp^2 + q_\perp^2 + 2 k_\perp q_\perp \cos (\phi_1 -\phi_2) }   \nonumber\\
 &\sim&  \int_{ k_\perp,q_\perp >\frac{\rho}{L}}    d k_\perp  d q_\perp   \frac{ 1}{ k_\perp^2 + q_\perp^2 }  
  \nonumber\\
  &= & \frac{\pi}{2} \ln \left( \frac{\rho}{L} \right) 
\ea
Note that in the second line we have neglected the integrals over $\phi_1$ and  $\phi_2$, 
the angular directions of  $\bfk_{\bot}$ and $\bfq_{\bot}$, which do not change the IR behavior of the integral. Similarly, for the second term in big bracket in Eq. (\ref{var-asym3}) we obtain
\ba
\label{asym2}
 \int_{ k_\perp,q_\perp >\frac{\rho}{L}}    d k_\perp  d q_\perp   \frac{ k_\perp q_\perp }{ (k_\perp^2 + q_\perp^2 )^2}  \simeq \frac{\pi}{2} \ln \left( \frac{\rho}{L} \right)  \, .
\ea
Combining the asymptotic results Eqs. (\ref{asym1}) and (\ref{asym2}) we obtain 
\ba
\label{var-asym4}
\Delta \langle \calR(\bfx)^2 \rangle_{{ \rm asym.}} \sim  -\frac{\epsilon }{16 \pi^3} 
\Big( \frac{H^{2}}{\dot \phi} \Big)^{2}  \ln \left( \frac{\rho}{L} \right)  \, .
\ea

Indeed, the above result can be confirmed by taking the integral in Eq. (\ref{var-asym3})  analytically.
With some efforts one can take the integral over $k_2$ and $q_2$ in Eq. (\ref{var-asym3}),  obtaining  
\ba
\Delta \langle \calR(\bfx)^2 \rangle_{{\rm asym.}}&=&
-\frac{\pi^3  \epsilon}{(2\pi)^6} \Big( \frac{H^{2}}{\dot \phi} \Big)^{2}  \int dk_1 dq_1  \frac{e^{i(k_1+q_1)\rho}}{(k_1+q_1)^2} \nn\\ &&
\times \Big(\mathrm{sgn} (k_1+q_1)+\mathrm{sgn}(k_1)\Big)\Big (\mathrm{sgn}(q_1)+\mathrm{sgn}(2k_1+q_1)\Big ) \\ 
&=&\frac{-\pi^3  \epsilon}{(2\pi)^6} \Big( \frac{H^{2}}{\dot \phi} \Big)^{2} \int dk dq_1  \frac{e^{ik\rho} }{k^2} %\nonumber \\ \times 
 \Big (\mathrm{sgn}(k)+\mathrm{sgn}(k-q_1)\Big )\Big (\mathrm{sgn}(q_1)+\mathrm{sgn}(2k-q_1)\Big ) \, ,
 \nn
\ea
Taking the integral and being careful on sign functions gives the following result for the asymmetric variance 
\ba
\label{var-asym5}
\Delta \langle \calR(\bfx)^2 \rangle_{{ \rm asym.}}  &=&-\frac{\pi^3  \epsilon}{(2\pi)^6} ( \frac{H^{2}}{\dot \phi} )^{2}
\times 4(-2\pi^3)
\mathrm{Re} \Big (\int_{\rho/L}\frac{dk}{k}\exp (ik)\Big ) \nonumber \\ 
&=&-\frac{\epsilon }{8\pi^3}  \left( \frac{H^{2}}{\dot \phi}  \right)^{2} \ln (\rho/L) \nonumber\\
&=& -\frac{\epsilon}{2 \pi} \calP_\calR^{(0)} \ln (\rho/L)  \, .
\ea
Interestingly, we see that our approximate result Eq. (\ref{var-asym4}) is well consistent (up to a factor of 2) with  the exact result in Eq. (\ref{var-asym5}). 

In the next section we estimate the observational bounds on the model parameters by comparing 
Eq. (\ref{var-asym5}) with the Planck data. 

%%%%%%%%%%%%%%%%%%%%%%%%%%%%%%%%%%%%%%%%%%%%%%%%%%
\begin{figure}[t]
	\begin{center}
	\subfloat[]{
	\includegraphics[scale=0.62]{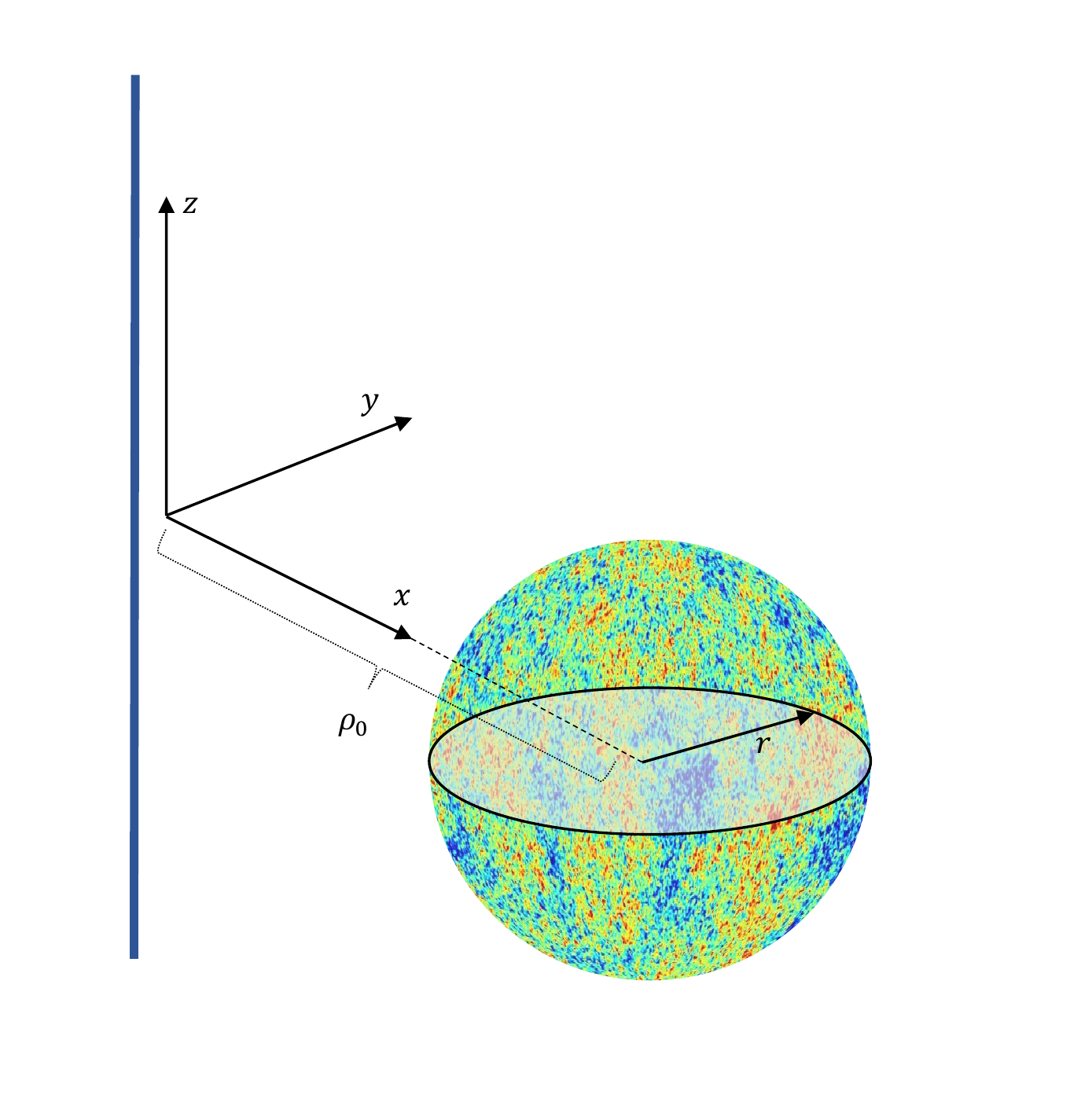}}
\subfloat[]{
	\includegraphics[scale=0.31]{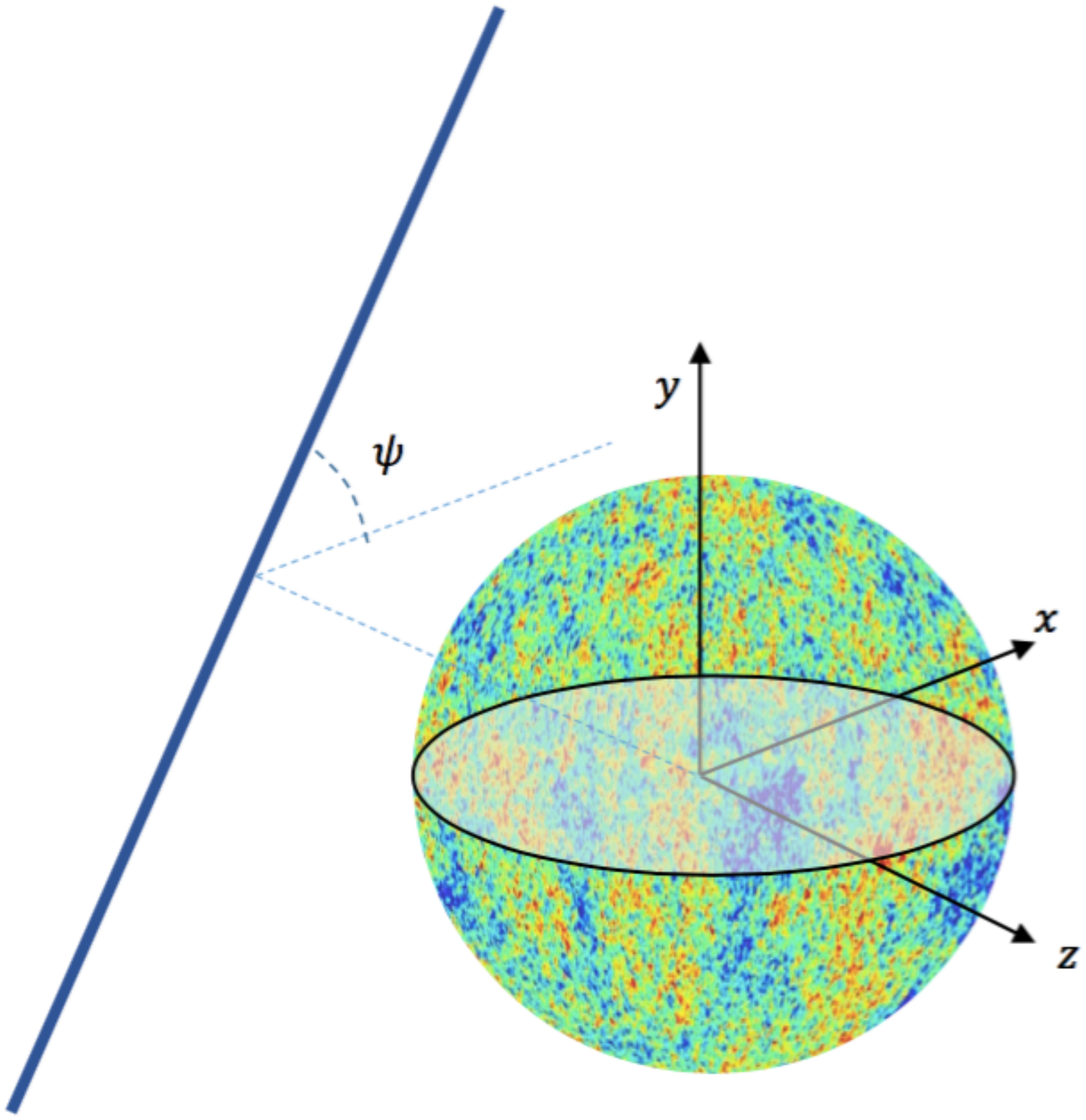} }
\end{center}
\caption{(a): The original coordinate system in which the string is orientated along the $\hat { z}$ direction 
where  we have calculated the corrections in power spectrum. (b): The new coordinate in which we perform numerical analysis for variance.  The direction of  dipole asymmetry  is towards the $-\hat{ z}$ direction. In galactic coordinate we have $\hat{ z}=(l,b)=(44^{\circ},22^{\circ})$ and $\hat{ x}=(44^{\circ},-68^{\circ})$.}
\label{fig1}
\end{figure}
%%%%%%%%%%%%%%%%%%%%%%%%%%%%%%%%%%%%%%%%%%%%%%%%%%

%%%%%%%%%%%%%%%%%%%%%%%%%%%%%%%%%%%%%%%%%%%%%%%%%%%
\section{Numerical results and comparison with observations}
\label{numeric}

Having obtained the analytical estimate for the variance of curvature perturbation in Eq. (\ref{var-asym5}) in this section we look for the constraints on the model parameters by comparing our analytical result for the variance with the Planck data.

%%%%%%%%%%%%%%%%%%%%%%%%%%%%%%%%%%%%%%%%%%%%%%%%%%%
\subsection{Variance of the TT map}

In \cite{Akrami:2014eta}, the authors have constructed a map of variance  out of the TT map of the Planck data. They have obtained the best fit values for the direction of the variance asymmetry as well as the multipole moments of the variance map. In their analysis  they have assumed the $SO(2)$ symmetry for the map of the variance  of  temperature fluctuations. However, in our model by locating an infinite cosmic string near our Hubble patch during inflation, we have spontaneously broken all rotational symmetries as well as the two-dimensional translational group on the plane perpendicular to string. Nevertheless, in order to constrain our model's parameters with the actual data and especially with the results  of \cite{Akrami:2014eta}, we average over the direction of cosmic string to resume the $SO(2)$ symmetry and thereby to compute the multipole moments of anisotropies in the  map of variance. 

Since we average over the orientation of sting in a two-dimensional plane, the preferred direction is now the direction perpendicular to this plane. We change the coordinate system such that now the third axis is perpendicular to this plane. One particular realization of string is identified by the angle $\psi$ measuring 
the angle of string with the new  $\hat { x}$ direction. For a schematic view of the orientation of string in the old and the new coordinate system look at Fig. \ref{fig1}. 

%%%%%%%%%%%%%%%%%%%%%%%%%%%%%%%%%%%%%%%%%%%%%%%%%%%
\begin{figure}[t]
	\begin{center}
	\includegraphics[scale=0.2]{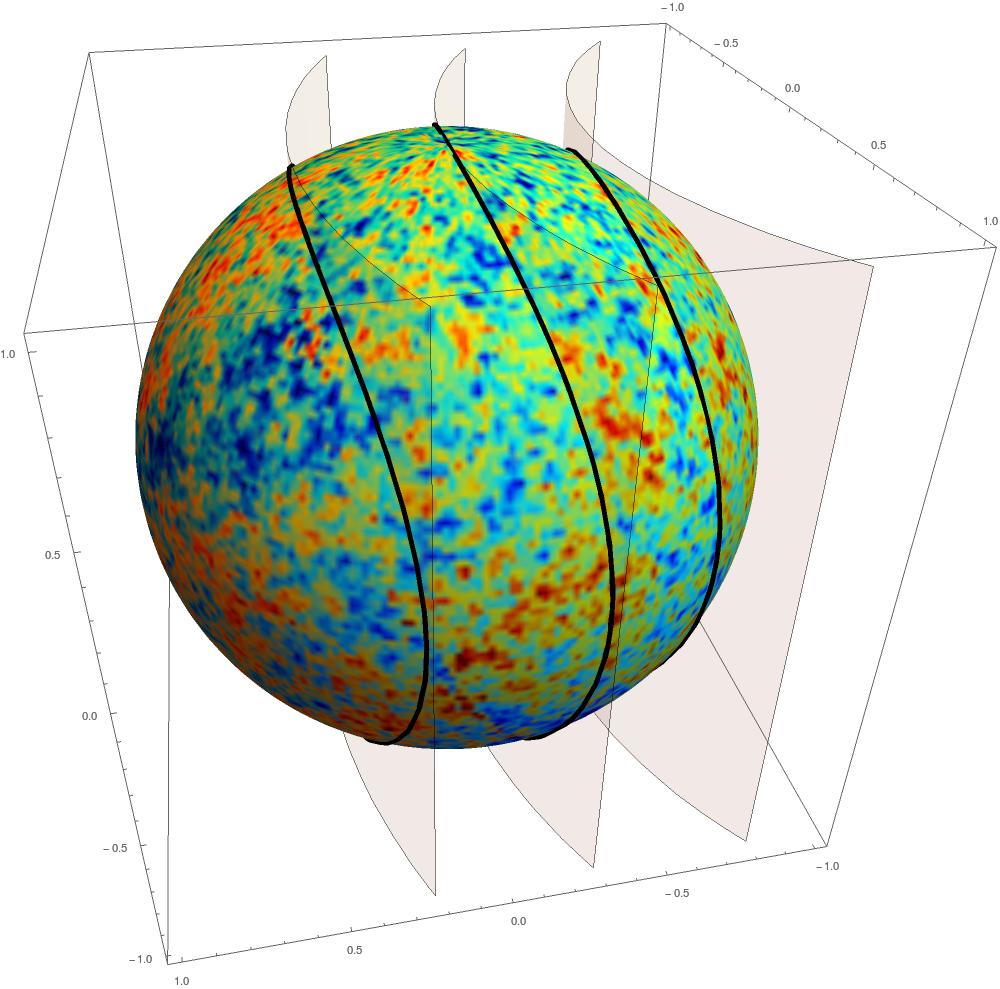}
\hspace{1.5cm}
	\includegraphics[scale=0.2]{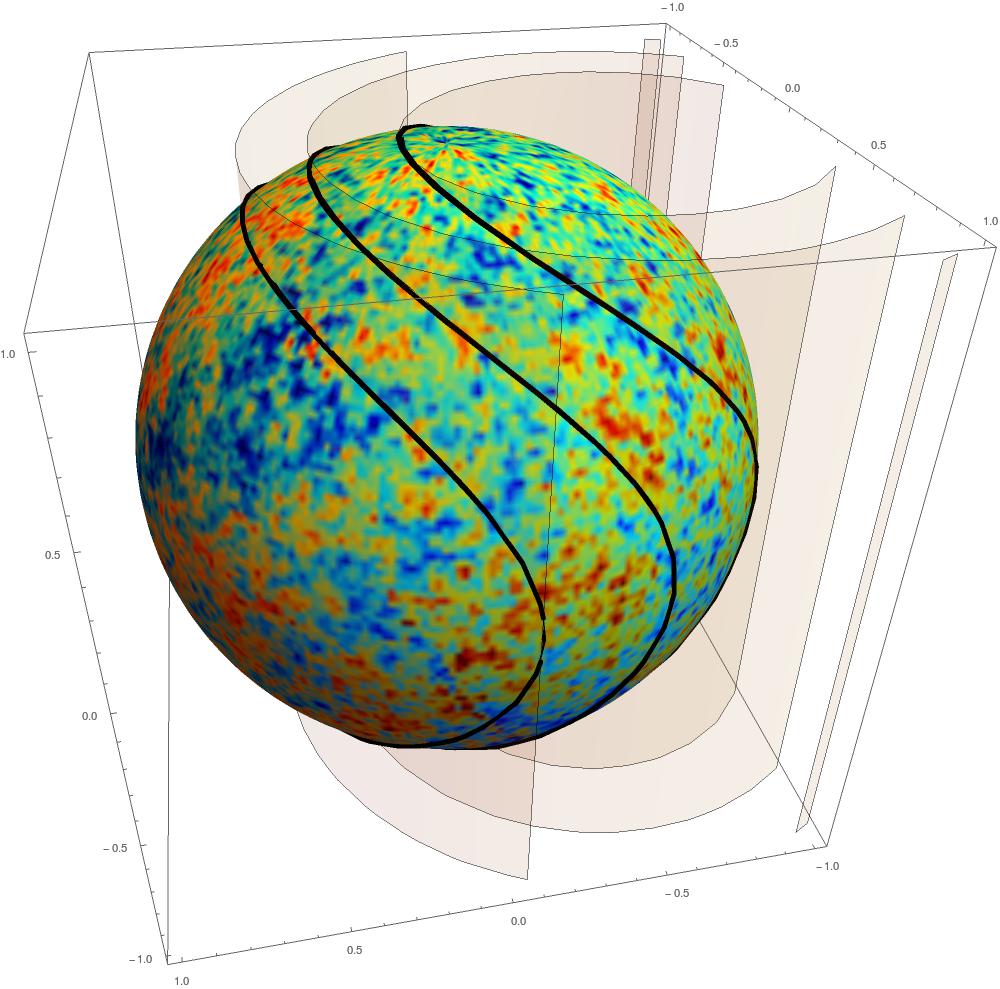} 
\end{center}
\caption{The curves of equal variance on CMB sphere induced by cosmic string,
left: $\kappa =0.5$, right: $\kappa =2$.  These curves are obtained by the intersection of hypersurfaces $\rho={\rm constant }$  with the CMB sphere. %where $\rho$ is the distance of string from the center of CMB sphere. 
}
\label{eq-var-fig}
\end{figure}
%%%%%%%%%%%%%%%%%%%%%%%%%%%%%%%%%%%%%%%%%%%%%%%%%%%
%%%%%%%%%%%%%%%%%%%%%%%%%%%%%%%%%%%%%%%%%%%%%%%%%%%
\begin{figure}[h]
\vspace{-1.5cm}
\begin{center}
\includegraphics[scale=0.4]{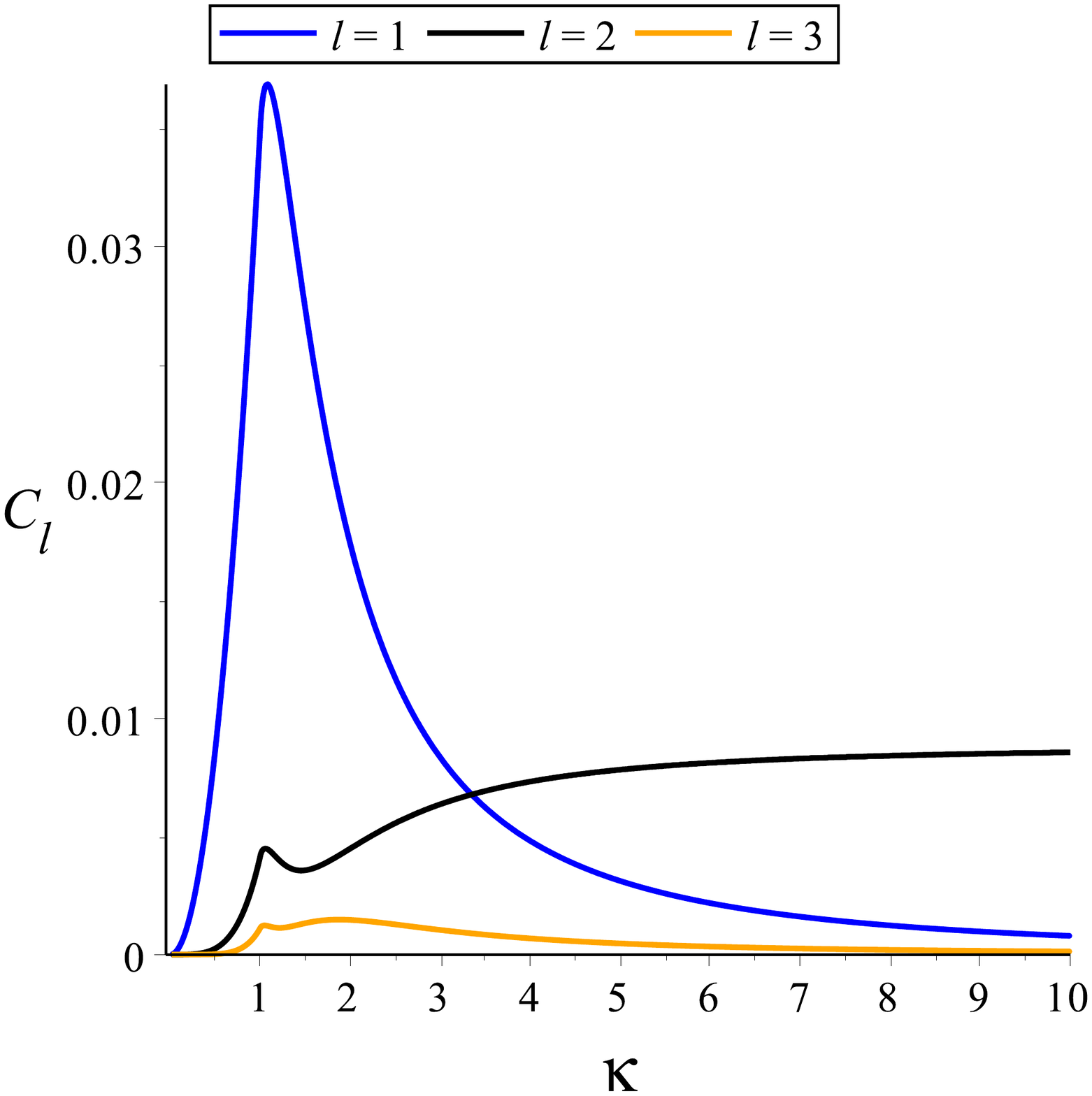}
\end{center}
\vspace{-2cm}
\caption{ Angular power spectrum of the variance as a function of $\kappa $ 
for dipole, quadrupole and octupole 
with the normalization   $\epsilon=1$. }
\label{Cl-fig}
\end{figure}
%%%%%%%%%%%%%%%%%%%%%%%%%%%%%%%%%%%%%%%%%%%%%%%%%%%

In the new coordinate system, the anisotropic correction in 
variance of curvature perturbation from Eq. (\ref{var-asym5}) (after removing the constant isotropic piece)  is given by 
\ba
\label{var}
{\rm Var}(\theta,\phi | \epsilon,\kappa,\psi)=-\dfrac{\epsilon}{4\pi}\ln \Big(1+\kappa^2\sin^2\theta\sin^2(\phi-\psi)+\kappa^2\cos^2\theta+2\kappa\cos\theta\Big) \, ,
\ea
in which we have defined  $\kappa \equiv {r}/{\rho_0}$ where $\rho_0$ is the distance between string and the center of  CMB sphere and $r$ is the comoving radius of CMB sphere  as shown in Fig. \ref{fig1}. In Fig. \ref{eq-var-fig} we have plotted the curves of constant variance on the CMB sphere. These curves are obtained by intersecting the hypersurfaces of constant $\rho$ 
with the CMB sphere.

Now we need to average over the string direction to compute an angular spectrum for the map of variance. In order to do this, firstly we compute the $a_{lm}$ multipoles associated with \eqref{var}:
\ba
a_{lm}\equiv \int d\Omega \hspace{1mm}Y_{lm}^*(\theta,\phi) Var(\theta,\phi) \, .
\ea
Summing over $m$ should give us a good measure for asymmetry. In other words, this summation removes the string direction, leaving the observer with the averaged angular power spectrum 
\ba
C_l\equiv \frac{1}{2l+1}\sum_{m}^{}|a_{lm}|^2 \, .
\ea
This is the quantity   computed in \cite{Akrami:2014eta} for both simulations and real data.  In Fig. \ref{Cl-fig} we have plotted $C_{l}$ for dipole, quadrupole and octupole as a function of $\kappa$.

%%%%%%%%%%%%%%%%%%%%%%%%%%%%%%%%%%%%%%%%%%%%%%%%%%%
\begin{figure}[h]
	\centering
	\includegraphics[scale=0.29]{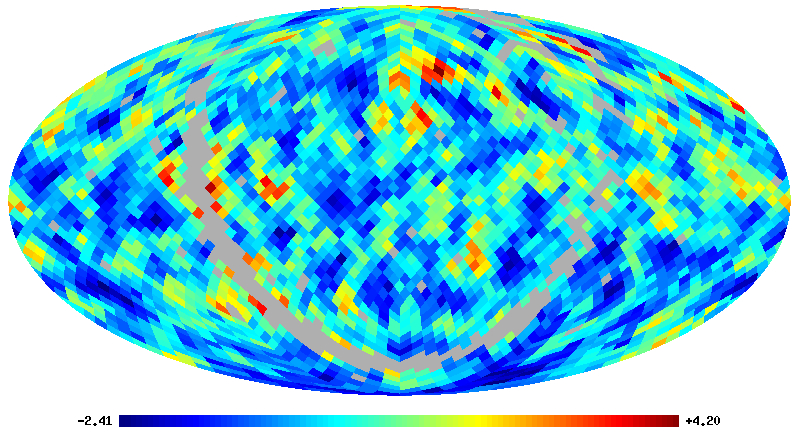}
	\includegraphics[scale=0.29]{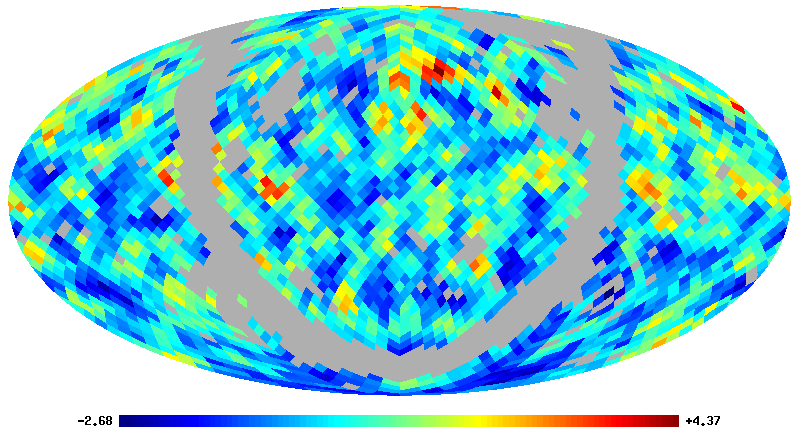}
	\caption{{Left:} NILC masked variance map, {right:} SMICA masked variance map. They are masked by their individual masks and rotated such that the vertical direction lies in the direction of dipole variance asymmetry, namely $(\ell,b)= (224,-22)$ in galactic coordinates.}
	\label{cmbvar}
\end{figure}
%%%%%%%%%%%%%%%%%%%%%%%%%%%%%%%%%%%%%%%%%%%%%%%%%%%

In order to find the best fit values of our free parameters, we have rotated the Planck's map of variance of fluctuations such that the direction of the reported dipole asymmetry  (in spherical coordinate) is along  the $-\hat{ z}$ direction, see Figus. \ref{fig1} and \ref{cmbvar}. The angle of string with respect to $\hat{ x}$ in new coordinate system is represented by $\psi$.

\begin{figure}[t]
	\centering
		\includegraphics[scale=0.7]{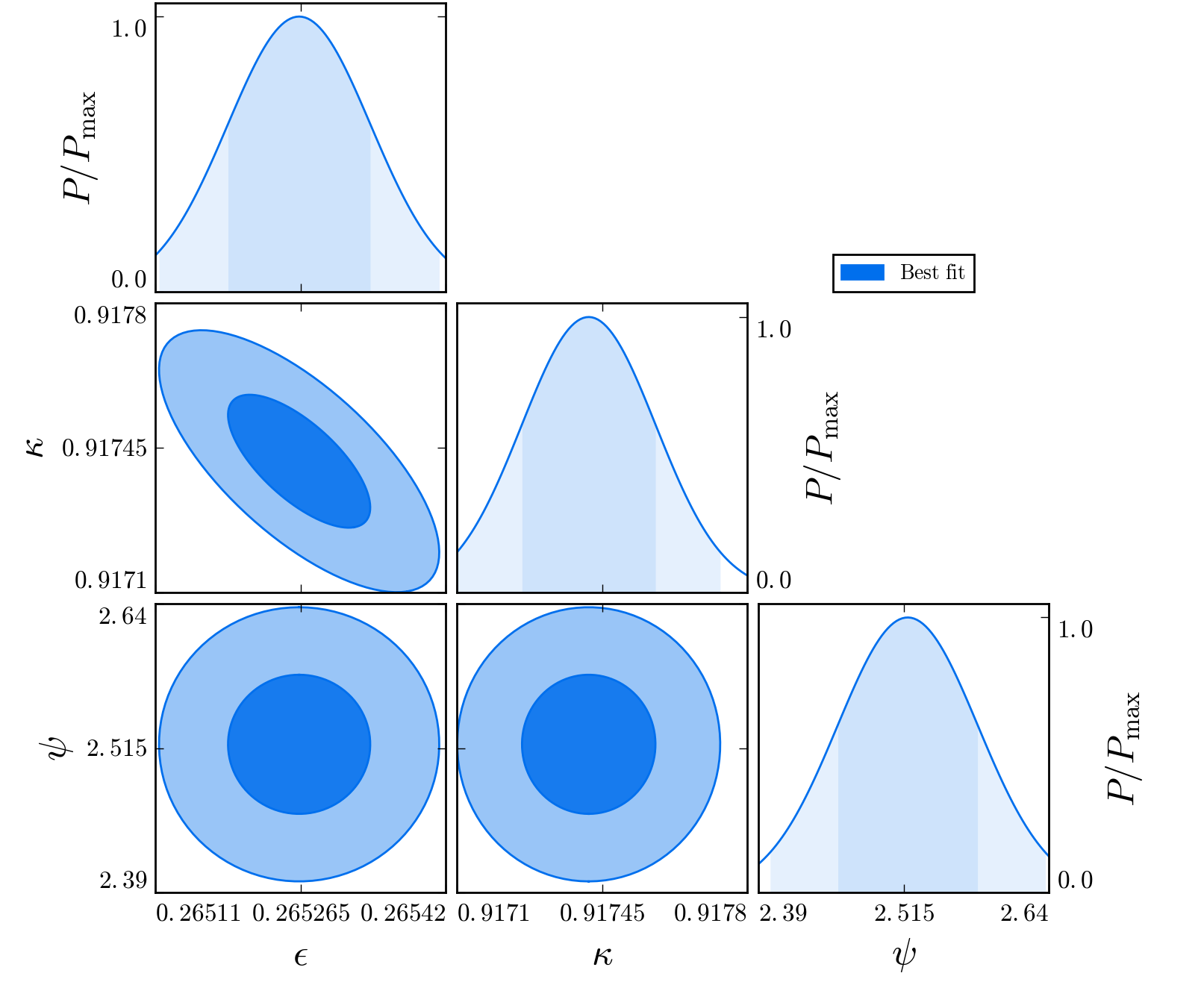}
			\caption{The three parameters Fisher analysis with the  best values $(\epsilon,\kappa,\psi)=(0.265,0.917,2.518)$.}
\label{Fisher}
\end{figure}

Now, by means of a likelihood analysis and comparing  $a_{lm}$ with the data,  one can find the best fit values for $\epsilon$, $\kappa$ and $\psi$.  For this purpose, we used all
CMB component separation algorithms, namely \textit{Commander ruler}, \textit{NILC}, \textit{SMICA}, and \textit{SEVEM} maps of Planck DR-2 intensity maps and the corresponding masks for each component \cite{Adam:2015tpy}. We extracted variance map out of these maps by calculating the variance of the TT fluctuations over $6^{\circ}$ circles on the CMB sphere  and applied the corresponding mask for each map. Afterwards we used those disks that had more than 90 percent unmasked pixels to construct a variance map and ignored other disks. Finally, we computed $a_{lm}$ for all of our variance maps and tried to maximize the likelihood function over the parametric space.  Since $a_{lm}$ decays rapidly for large $\ell$, at the first level of analysis, we have limited ourselves to  $\ell \le 10$. Then we  have found the best fit of parameters according to a pre-analysis obtained from the $C_\ell$ analysis like in \cite{Akrami:2014eta}. We found that this model can not be a good fit to all $C_\ell$ (or $a_{\ell m}$) of the variance map. The reason is that  if we try to fit high (say $\ell>4$) multipoles simultaneously, then we miss lower multipoles and the model ceases to be a good fit to $C_\ell$ values for $\ell \leq 3$ . Also according to \cite{Akrami:2014eta},  lower multipoles have larger confidence level compared to high $\ell$ values, hence 
when looking for the best fit values for our model's parameters, we coarse grain the variance map by looking at $\ell<4$ multipoles. The best fit values extracted out of this procedure for our parameters are $(\epsilon,\kappa,\psi)=(0.265,0.917,2.518)$, see Fig. \ref{Fisher}. 

There are two important points to mention. First,  we see that this best fit value, $\epsilon = 0.265$, 
obtained  from the dipolar asymmetry is an order of magnitude weaker than the constraint 
$\epsilon \lesssim 10^{-2}$ obtained from the quadrupolar anisotropy.  Second, the best fit value $\kappa =0.917$ corresponds to the configuration in which the string is very close to CMB sphere. In realistic situation, it requires fine-tuning so one expects $\kappa$ to be somewhat different than unity.

%%%%%%%%%%%%%%%%%%%%%%%%%%%%%%%%%%%%%%%%%%%%%%%%%%%
%%%%%%%%%%%%%%%%%%%%%%%%%%%%%%%%%%%%%%%%%%%%%%%%%%%
\begin{figure}[h]
	\begin{center}
	%\subfloat[]{
	\includegraphics[scale=0.065]{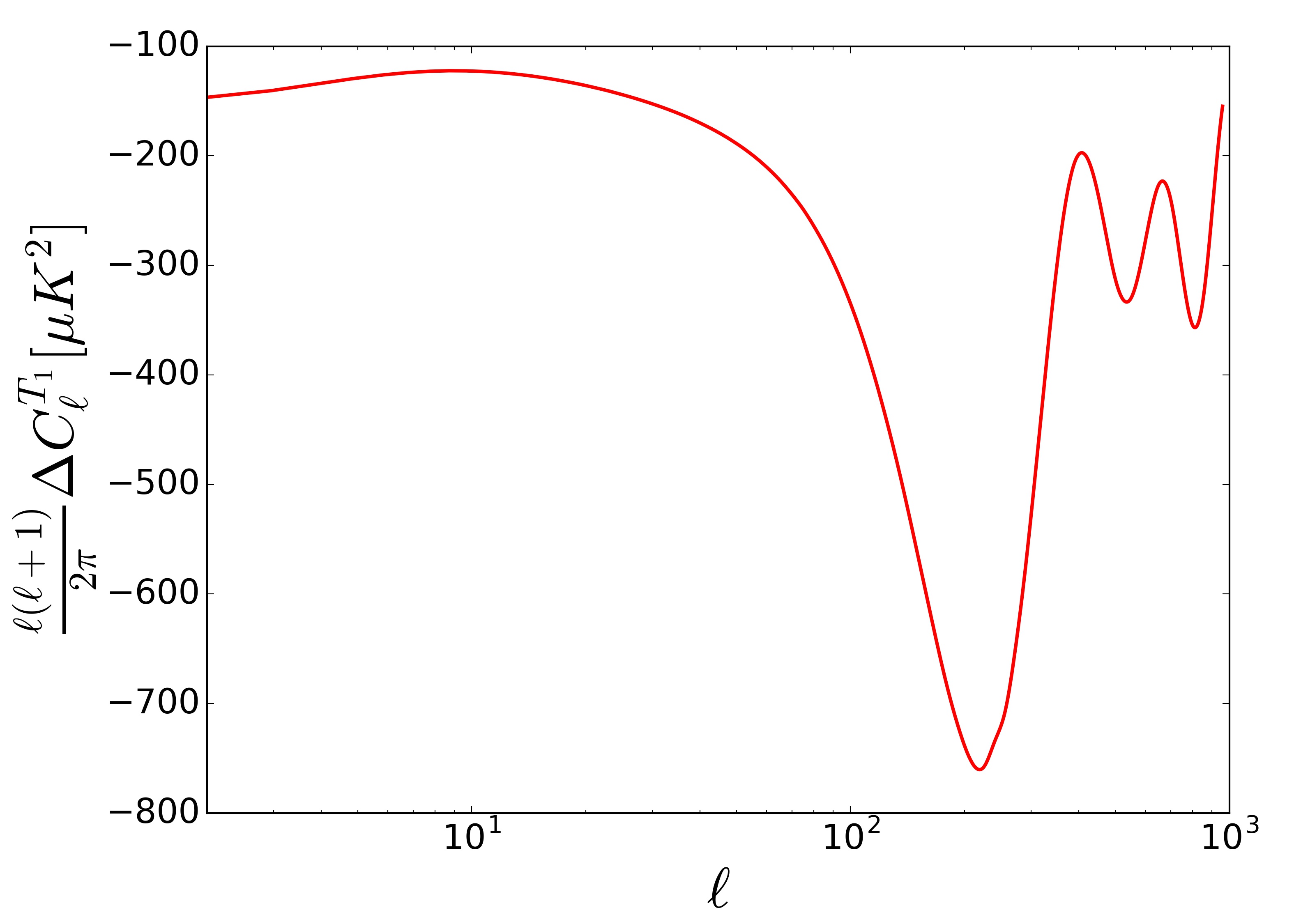}
%}
%\subfloat[]{
\hspace{0.7cm}
	\includegraphics[scale=0.065]{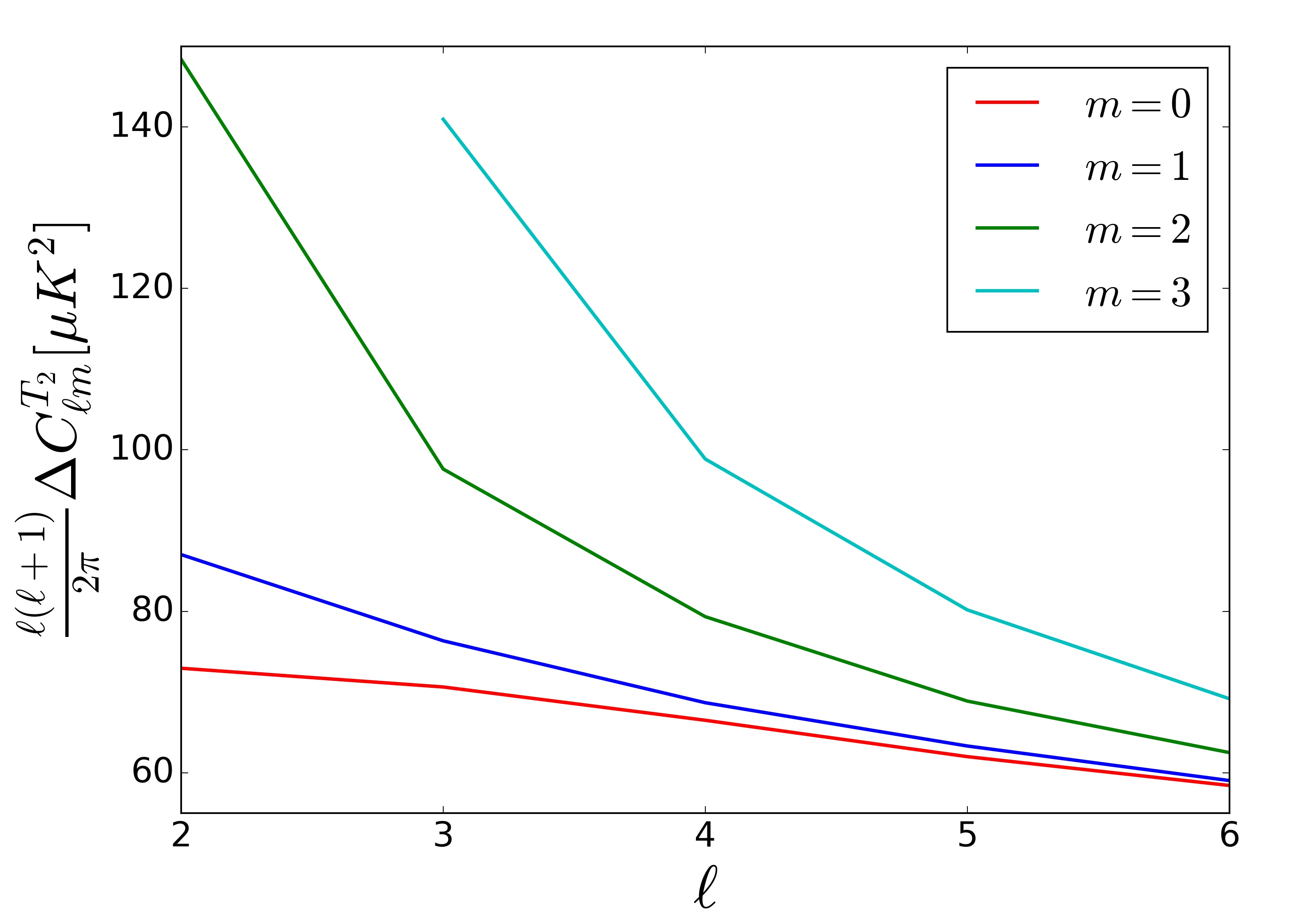}
\end{center}
	\begin{center}
		%\subfloat[]{
	\includegraphics[scale=0.065]{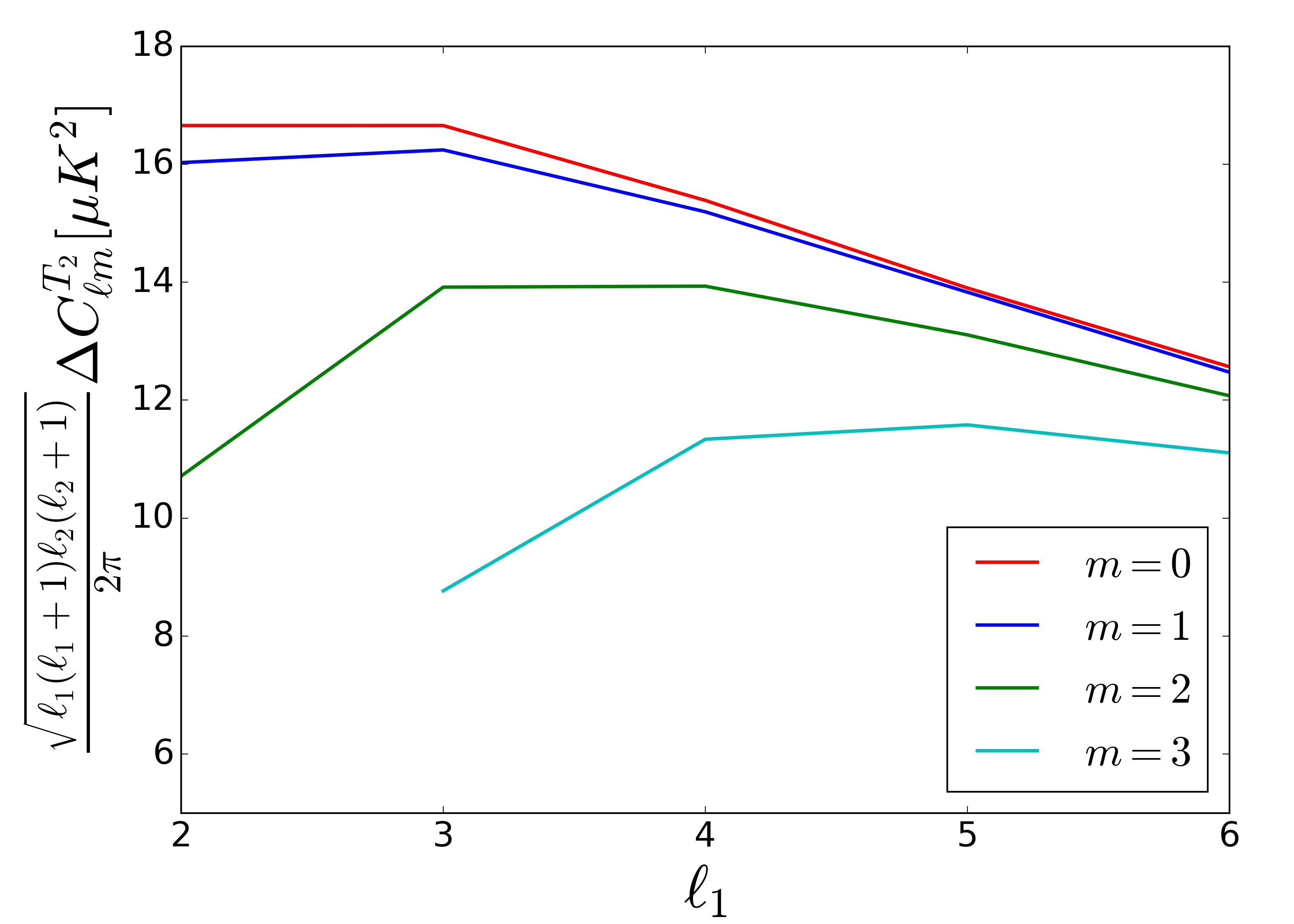}
	%}
%\subfloat[]{
\hspace{0.7cm}
	\includegraphics[scale=0.065]{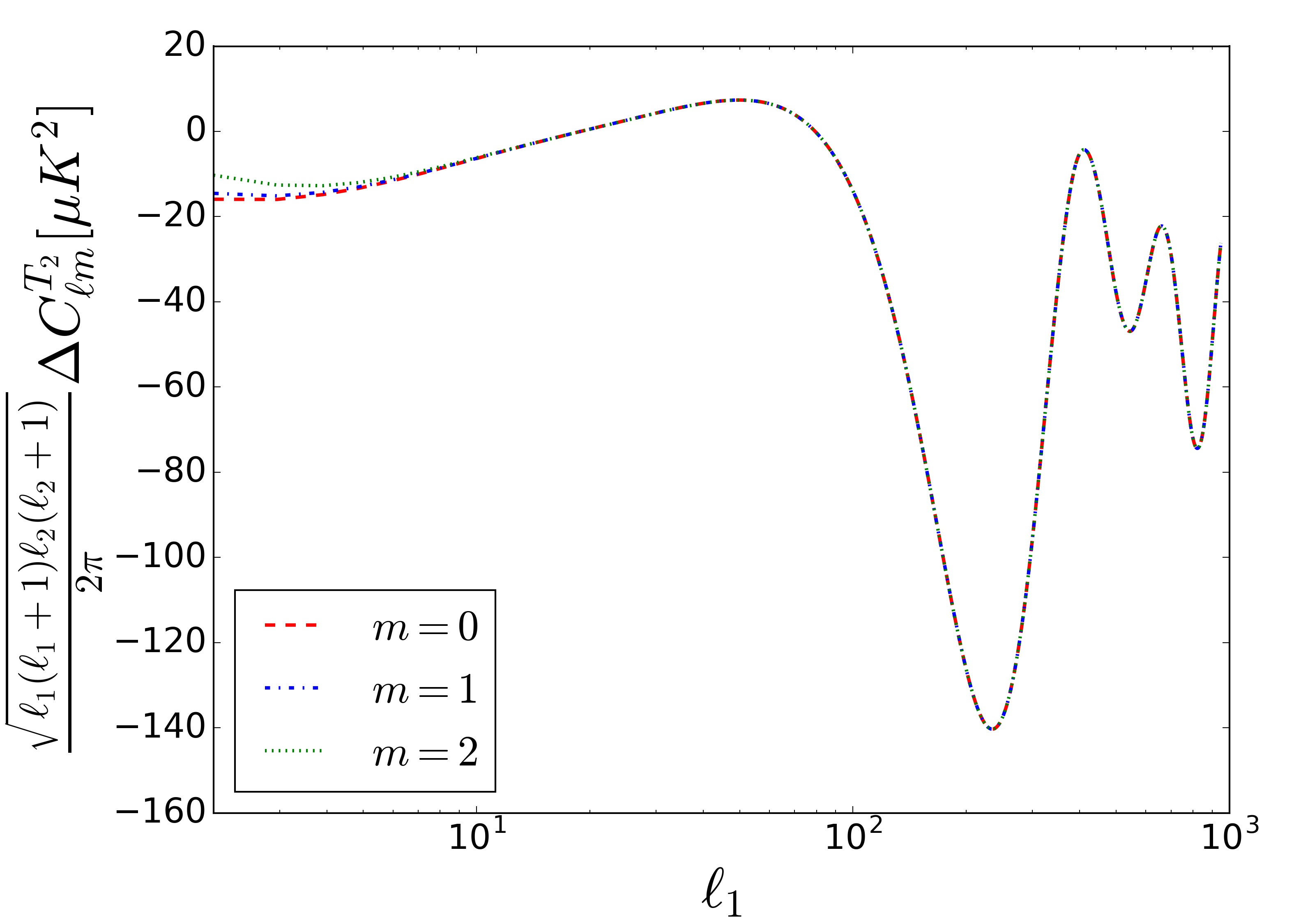}
	%}
	\end{center}
	\caption{ Some elements of the first (homogenous) and second (inhomogeneous) parts of the power spectrum  Eq. (\ref{Delta-Wise}) for the angular power spectrum matrix, $C_{(l,m)(l',m')}$, evaluated for the best fit values found by variance analysis in Fig. \ref{numeric}. Top left: the diagonal part of the first contribution with the sum over $m$.  Top right: the diagonal part of the second contribution  evaluated for different $m$. Since the computational cost of calculating this part is very high, we did not sum over all $m$.
	Lower left: the $l_2=l_1+2$ elements of second part of the angular power, evaluated for different $m$.  Lower right:  the $l_2=l_1+2$ element (the only non zero off diagonal element) of the first contribution for different $m$.}
	\label{Cl}
\end{figure}

%%%%%%%%%%%%%%%%%%%%%%%%%%%%%%%%%%%%%%%%%%%%%%%%%%%
\subsection{Angular spectrum of TT map}

Here we perform the analysis of  CMB angular two point function.
 
Computing the angular two point function of the TT map with the  primordial  curvature  power spectrum 
Eq. (\ref{Delta-Wise}) is straightforward. The details of the formulas are reported in Appendix  \ref{appA}. It is useful to decompose the primordial spectrum into two parts as represented in Eq. \eqref{bipo}. The first part does not violate the translational invariance and as mentioned before  is simply a quadrupole term, while the second part breaks the translational invariance. These two parts have different contribution into the angular power spectrum, hence in the following we compute and plot each contribution separately. %denoting them by $I$ and $II$. 

The results shown in Fig. \ref{Cl} are plotted for the best fit values found in Fig. \ref{numeric}, namely 
$(\epsilon,\kappa)=(0.265,0.917)$. The second contribution which violates translation invariance rapidly decays for large $\ell$s, as a result the first contribution  which is homogeneous  dominates over the non-homogeneous part for  $\ell>3$ for diagonal elements. We also observe that the $\ell_2=\ell_1$ part of the second contribution is much bigger than its off-diagonal $l_2=l_1+2$ elements. Computing the second contribution is numerically too expensive so we calculated only its low multipole elements.\footnote{We have to keep in mind that the theoretical value for the angular power spectrum depends on coordinate we choose, due to the lack of rotational symmetry. Consequently, one can not directly compare the diagonal $C_{ll'mm'}$ terms found here with the actual $C_l$ plots of Planck's data. Nevertheless, what we plot here should give a rough picture of how $C_l$ would look like if we rotate our TT map in order to match the coordinate in Planck's map and if we average properly over $m$ and $m'$. }

%%%%%%%%%%%%%%%%%%%%%%%%%%%%%%%%%%%%%%%%%%%%%%%%%%%

\section{Discussions}

In this work we have looked for the imprints of a primordial cosmic string during inflation in generating 
statistical anisotropy and power asymmetry. 
The question of looking for the effects of cosmic strings in early universe is very well motivated. Cosmic strings can be generated  from  a $U(1) $ symmetry breaking  during inflation. Alternatively,  they can be the F- and D- strings of superstring theory. In either ways, constraining the tension of cosmic string directly constrains the mass scale of the corresponding underlying theories responsible for the formation of cosmic strings.

The contribution of cosmic string in curvature power spectrum has two distinct parts. The first part is homogenous and has the form of quadrupolar statistical anisotropy. Comparing with the Planck constraints on the amplitude of quadrupolar anisotropy we obtain the upper bound $G \mu \lesssim 10^{-2}$ so the energy scale 
of the underlying theories generating  cosmic string can not be significantly higher than the GUT scale. 

The second contribution of cosmic string in curvature power spectrum breaks the translation invariance in the plane perpendicular to string. This contributes to asymmetry in variance of curvature perturbations. The resulting 
constraint on the tension of cosmic string $G \mu \sim 10^{-1}$ is about an order of magnitude weaker than the constraint from the quadrupolar anisotropy. 

We have calculated the contributions of the above mentioned two terms in CMB angular power spectrum. Because the isotropy and the homogeneities are broken, we will have off-diagonal contributions in angular power spectrum. 
The contribution of the inhomogeneous part rapidly falls off with $\ell$ for both diagonal and off-diagonal part. The non-trivial scale dependence of power asymmetry from the inhomogeneous term is a good news, as the observed dipole asymmetry in CMB maps suggest a rapid fall off for the power asymmetry in small scale.  Having said this, a dedicated data analysis is required to investigate the full effects of strings on CMB temperature and polarization  maps. 

In our analysis we have considered the simple picture of an infinite  string  straight. In a realistic situation, one may encounter a network of cosmic strings during inflation. So it is an interesting question what would be the imprints of a network of cosmic strings with a mix of loops and long strings on inflationary power spectrum. During inflation the strings are diluted quickly so if one waits for few e-folds then our picture of a long straight string  is well justified. However, during the short transient regime when the strings are being diluted, the imprints of a network of cosmic strings in inflationary power spectrum would be much more complicated than our results. It may be an interesting question to look for the transient effects of a network of cosmic string during early stage of inflation and to see whether a network of cosmic strings can address the anomalies 
on CMB maps. \\\vspace{0.2cm}

%%%%%%%%%%%%%%%%%%%%%%%%%%%%%%%%%%%%%%%%%%%%%%%%%%%%%

{\bf Acknowledgments:}  
We thank  Yashar Akrami, Jaume Garriga, Martin Kunz and Alessio Notari  for useful discussions. H. F. would like to thank the University of Barcelona for hospitality where this work was at its final stage. 
The computations were performed at University of Geneva on the Baobab cluster.

\appendix 

%%%%%%%%%%%%%%%%%%%%%%%%%%%%%%%%%%%%%%%%%%%%%%%%%%%
\section{Angular power spectrum}
\label{appA}
The relation between the primordial  curvature perturbations and angular fluctuations of the CMB is given by 
\be
a_{lm}=4\pi i^l \int \frac{d^3k}{(2\pi)^3}\Delta_l (k)\calR_{\bk}Y_{lm}(\hat{k}) \, ,
\ee
where $\calR_{\bfk}$ is the curvature perturbations of a particular mode. 

As discussed in the main text the corrections from cosmic string in power spectrum, given in Eq. (\ref{Delta-Wise}), have  two distinct parts: 
\be
\label{bipo}
\Delta \la {\cal R}_{\bk}{\cal R}^*_{\bq}\ra={\cal F}_1(\bk)\delta^3(\bk-\bq)+{\cal F}_2(\bk_\bot,\bq_\bot)\delta(k_z-q_z) \, .
\ee
The term ${\cal F}_1$ violates the isotropy but not the homogeneity, while ${\cal F}_2$ violates both isotropy and homogeneity. These functions are given by the following formulas: 
\ba
{\cal F}_1(\bk)&=&12\pi^5 \epsilon {\cal P}_\calR^{(0)} \dfrac{\sin^2\theta}{k^3}\\ \nn
{\cal F}_2(\bk_\bot,\bq_\bot)&=&-(2\pi)^4\epsilon {\cal P}_\calR^{(0)} \dfrac{ \exp \Big(i(k_1-q_1)\rho\Big)}{k^3q^3}\Big(\dfrac{k^2+q^2+kq}{k+q}\Big)
\\ \nn
&&\dfrac{1}{(\bk_\bot-\bq_\bot)^2}\Big \lbrace \frac{1}{2} \bk_\bot .\bq_\bot-\frac{1}{(\bk_\bot-\bq_\bot)^2}\bk_\bot .(\bk_\bot-\bq_\bot) \bq_\bot .(\bk_\bot-\bq_\bot)\Big \rbrace \, .
\ea

The matrix elements of the TT anisotropies are
\footnote{Note that $a_{lm}$ depends on the coordinate system we work with as shown in  Fig. \ref{fig1}. }
\be
C^{TT}_{(l,m)(l'm')}=\la a_{lm}a^*_{l'm'}\ra=(4\pi)^2 i^{l-l'}\int \frac{d^3k d^3q}{(2\pi)^6}\Delta_l(k)\Delta_{l'}(q)Y_{lm}(\hat{k})Y^*_{l'm'}(\hat{q})\la {\cal R}_{\bk}{\cal R}^*_{\bq}\ra  \, .
\ee
We separate the matrix elements due to different terms in \eqref{bipo}. The first piece contributes as  
\ba
\label{1piece}
C^{I}_{(l,m)(l'm')}=(4\pi)^2 i^{l-l'}\int \frac{d^3k}{(2\pi)^6}\Delta_l(k)\Delta_{l'}(k){\cal F}_1(\bk)Y_{lm}(\hat{k})Y^*_{l'm'}(\hat{k}) \, .
\ea
Hereafter the following convention for spherical harmonics functions is being used:
\be
Y_{lm}(\theta,\phi)=\sqrt{\frac{(2l+1)}{4\pi}\frac{(l-m)!}{(l+m)!}}P_l^m(\cos \theta)\exp(im\phi) \, .
\ee
Correspondingly, Eq.  \eqref{1piece} simplifies to
\ba
\label{c1}
C^{I}_{(l,m)(l'm')}&=&\delta_{mm'}\sqrt{(2l+1)\dfrac{(l-m)!}{(l+m)!}}
\sqrt{(2l'+1)\dfrac{(l'-m)!}{(l'+m)!}}(4\pi) i^{l-l'}\\ \nn
&&\times \int \dfrac{k^2dk \sin \theta d\theta }{(2\pi)^5}\Delta_l(k)\Delta_{l'}(k){\cal F}_1(k,\theta)P_l^m(\cos \theta)P_{l'}^{m*}(\cos \theta) \, .
\ea
As for the second term  we have: 
\ba
\label{c2}
C^{II}_{(l,m)(l'm')}&=&(4\pi)^2 i^{l-l'}\int \dfrac{k^2 dk d\phi  \sin \theta d\theta}{(2\pi)^6}QdQd\phi'\Delta_l(k)\Delta_{l'}(\sqrt{Q^2+k^2\cos^2 \theta})\\ \nn
&&\times Y_{lm}(\cos \theta, \phi)Y^*_{l'm'}(\frac{k\cos \theta}{\sqrt{Q^2+k^2\cos^2 \theta}},\phi'){\cal F}_2(\bk_\bot,\bq_\bot) \, .
\ea
These expressions are used in our analysis to calculate the angular  power spectrum in generating 
plots in Fig. \ref{Cl}.

%%%%%%%%%%%%%%%%%%%%%%%%%%%%%%%%%%%%%%%%%%%%%%%%%%%%%
{}

\end{document}